\UseRawInputEncoding
\documentclass[12pt, a4paper, twoside, titlepage, openright]{book}

\usepackage{float}
\usepackage{amsfonts}
\usepackage{dsfont}
\usepackage{amsmath}        
\usepackage{amssymb}        
\usepackage{latexsym} 
\usepackage[mathcal]{euscript}      
\usepackage{graphicx}                 % Packages to allow inclusion of graphics
\usepackage{color}                    % For creating coloured text and background
\usepackage{hyperref}                 % For creating hyperlinks in cross references
\usepackage{setspace}      
\usepackage{textcomp}			  % for onehalfspacing
\usepackage{marvosym} %\Mundus \Anglesign \Laserbeam 
\newcommand{\VL}[1]{}

\newcommand{\notonline}[1]{See printed copy of the thesis.}

\newcommand{\toquote}[2]{\begin{flushright}{\singlespacing\small\em #1 \\ ---#2}\end{flushright}}

\newcommand{\toabstract}[1]{\vspace{1.5cm}\begin{flushright}\rule{10.04cm}{0.05cm}\vspace{-0.35cm}\\
 \parbox{10cm}{\singlespacing\small #1}\\
\vspace{0.04cm} \rule{10.04cm}{0.05cm}\end{flushright}}

\newcommand{\bd}[1]{\textbf{#1}}
\newcommand{\ket}[1]{| #1 \rangle}
\newcommand{\bra}[1]{\langle #1 |}
\newcommand{\braket}[2]{\langle #1 | #2 \rangle}

\newcommand{\lra}{\longrightarrow}

\newcommand{\ie}{i.e. }
\newcommand{\eg}{e.g. }

\newtheorem{Th}{Theorem}[chapter]

\newtheorem{Cor}{Corollary}[chapter]
\newtheorem{Lem}{Lemma}[chapter]
\newtheorem{Rk}{Remark}[chapter]

\newtheorem{Post}{Postulate}

\newtheorem{Def}{Definition}[chapter]

\newtheorem{Ex}{Example}

\makeindex
\begin{document}

%-------------------------------------------------------------------------------------------
\frontmatter
%-------------------------------------------------------------------------------------------
\singlespace\pagestyle{plain}\pagenumbering{roman}

% A. TITLE
\title{
{\small ~}\\\vskip0.1cm
Universit\'e Aix-Marseille\\\vskip0.3cm
{\small ~}\\\vskip0.1cm
Habilitation \`a Diriger les Recherches\\\vskip3cm
{\bf \Huge Quantum walks, limits and transport equations}\vskip6cm}
\author{{\LARGE Giuseppe Di Molfetta}}
\date{{\small\em \copyright ~$14^{\textrm{th}}$ of Mai 2020}}
\maketitle

%B. ------------------------------------------------------------------------------------- E.

% B. ABSTRACT
\cleardoublepage
\section*{Abstract}\normalsize
 \addcontentsline{toc}{section}{Abstract}

This manuscript gathers and subsumes a long series of works on using QW to simulate transport phenomena. Quantum Walks (QWs) consist of single and isolated quantum systems, evolving in discrete or continuous time steps according to a causal, shift-invariant unitary evolution in discrete space. We start reminding some necessary fundamentals of linear algebra, including the definitions of Hilbert space, tensor state, the definition of linear operator and then we briefly present the principles of quantum mechanics on which this thesis is grounded. 
After having reviewed the literature of QWs and the main historical approaches to their study, we then move on to consider a new property of QWs, the plasticity. Plastic QWs are those ones admitting both continuous time-discrete space and continuous spacetime time limit. We show that such QWs can be used to quantum simulate a large class of physical phenomena described by transport equations. We investigate this new family of QWs in one and two spatial dimensions, showing that in two dimensions, the PDEs we can simulate are more general and include dispersive terms. 
We show that the above results do not need to rely on the grid and we prove that such QW-based quantum simulators can be defined on 2-complex simplicia, i.e. triangular lattices. Finally, we extend the above result to any arbitrary triangulation, proving that such QWs coincide in the continuous limit to a transport equation on a general curved surface, including the curved Dirac equation in 2+1 spacetime dimensions.

% C. ACKNOWLEDGMENTS
\clearpage
%\section*{Acknowledgements}\normalsize
 \addcontentsline{toc}{section}{Acknowledgements}  
 \textit{To the loves of my life, Vera and my sons.}

\clearpage

\section*{Collaborations and publications}\normalsize
 \addcontentsline{toc}{section}{Collaborations and publications}  

\begin{itemize}
\item Chapter \ref{chap:Continuous Limit1D} introduce the notion of Plasticity for QW on the infinite line and is only partially resulting from the paper \emph{A quantum walk with both a continuous-time limit and a continuous-spacetime limit}, co-authored with Pablo Arrighi (AMU), see \cite{molfetta2019quantum}.
\item Chapter \ref{chap:Continuous Limit2D} extends the Plastic QW to the grid and refers to the paper \emph{Continuous Time Limit of the DTQW in 2D+ 1 and Plasticity}, co-authored with Michael Manighalam (University of Boston) and recently accepted in QINP, see \cite{manighalam2020continuous}. 
\item Sections in Chapter \ref{chap:triangle} are results from the paper \emph{Dirac equation as a quantum walk over the honeycomb and triangular lattices}, co-authored with Ivan Marquez, Armando Perez (Universidad de Valencia) and Pablo Arrighi, see \cite{arrighi2018dirac}. 
\item Section \ref{sec:TQWcurved} generalises results obtained in the paper \emph{From curved spacetime to spacetime-dependent local unitaries over the honeycomb and triangular Quantum Walks}, co-authored with Ivan Marquez, Armando Perez and Pablo Arrighi, see \cite{arrighi2019curved}. 
\item Section \ref{sec:perspecitves} refers to a recent result obtained in the paper \emph{Dynamical Triangulation Induced by Quantum Walk}, co-authored with Quentin Aristote (ENS Paris) and Nathanael Eon, see \cite{aristote2020dynamical}. 
\item Section \ref{sec:perspecitves} refers to the paper \emph{Grover Search as a Naturally Occurring Phenomenon}, co-authored with Mathieu Roget (ENS Lyon), Stephan Guillet (ENS Lyon) and Pablo Arrighi, see \cite{roget2020grover}. 
\end{itemize}

\hspace{1cm}

\emph{Ivan Marquez} has been my PhD student until 2019 at AMU, co-supervised by Armando Perez (UV) and Pablo Arrighi (AMU). 
\emph{Nathanael Eon} is currently my PhD student at AMU co-supervised by Pablo Arrighi (AMU). \emph{Quentin Aristote}, \emph{Mathieu Roget} and \emph{Stephan Guillet} were all M2 students under my supervision.

\clearpage
 %-- proofread

% F. CONTENTS  
\addcontentsline{toc}{section}{Contents}
\tableofcontents
%\listoffigures
%\listoftables

%-------------------------------------------------------------------------------------------
\mainmatter
%-------------------------------------------------------------------------------------------
\singlespace\pagestyle{plain}\pagenumbering{arabic}

% word count 4012

\chapter{Introduction}\label{chap:introduction}

\toquote{“Science is not real progress until a new truth finds an environment ready to accept it.”}{Petr Kropotkin}

\clearpage

Classical computation is based on the abstract model of the Turing Machine, defined in 1936 by the English mathematician Alan Turing and subsequently reworked by John von Neumann in the 1940s. Later, at the beginning of the 80s, when the quantum theory was sufficiently advanced, it became clear that to efficiently simulate a quantum system a classical computer was no longer sufficient. Just in those years Richard Feynman argued that no classical Turing Machine could simulate a quantum system without an exponential slowdown of its performance. This is due to the space complexity needed to describe a quantum state. In fact such a state is typically described by a number of parameters that grows as $O(\exp(N))$ with $N$ size of the system and moreover, to simulate the temporal evolution of the system we also need a number of operations that grows exponentially with its size. Therefore, quantum theory lends itself to a combinatorial complexity of possible dynamic structures enormously large. It was Feynman himself in 1982 who proposed a solution to this problem. Quoting him: \textit{``Let the computer itself be built of quantum mechanical elements which obey quantum mechanical laws." (Feynman, 1982)}...a quantum machine has the capacity to contain an exponential explosion of information avoiding the use of an exponentially large amount of physical resources. It is important to notice that, although $N$ qubits can ``carry" a larger amount of (classical) information (thanks to quantum superposition), the Holevo's theorem proved that the amount of classical information that can be retrieved, i.e. accessed, can be only up to $N$ classical (non-quantum encoded) bits. \\

The idea that information can be stored in microscopic quantum states was an unprecedented challenge for scientists, since it opened the perspective of using quantum matter itself to make calculations. Unfortunately, the quantum coherence that would allow, \eg quantum algorithms to work, is very fragile. 

Before scientists could accept such a revolution, however, a few steps had to be taken in the world of atomic physics, of fundamental importance for the development of a quantum processor. The first steps in this direction was due to the physicist Hans Dehmelt (Nobel prize winner in 1989) who managed to isolate a single ion in a vacuum chamber and suspend it in the vacuum at a predetermined and controllable point. Then Zoller and Cirac realised that a single ion could act as a quantum gate, building the first quantum register with two distinct types of stored information depending on the physical characteristics of the ion. They had built the first qubit. The search for more stable and scalable quantum computer has never stopped. Today we can claim several technologies with up to 53 qubits, and capable of solving specific problems faster than classical computers. 

Feynman's intuition led to the birth of a new theoretical computer science, that had repercussions in the theory of computability, complexity and logic. For the first time in history, the sacrosanct ``hardware-independence'' principle was shaken. The physical hardware became fundamental once more, as it would condition the entire logic of the algorithm. We need to be very careful to the choice of the physical system, to its initialization and to its measurement. Whilst classical Computer Science has soon managed to prescind from the physical machine, in quantum computation, this seems to be difficult or even impossible. A close collaboration between the two communities is necessary, as the new questions raised lie precisely at the frontier between computer science and physics. In fact, if with the advent of modern computers we have seen a progressive separation of computer scientists from physicists and mathematicians, quantum computer science raises new questions intimately at the frontier between physics and theoretical computer science. 

It is in this increasingly vast and interdisciplinary boundary zone that as a student I took my first steps towards research and like me, hundreds of other young researchers. The increasing enthusiasm on these issues is not only related to the development of the first quantum computers, which in itself is attracting the interest of a large network of private and public actors. This enthusiasm has deeper roots: nowadays computing 
models conceived by computer scientists are becoming the new grammar for the study of natural processes. Information and its
processing, after having conquered quantum mechanics and thermodynamics, have now become central in the study of gravity or biology. In this revolution, computer scientists and physicists will work together, to develop a common language, thirty years after Feynman's words.

One of the fields in which such a symbiosis seems strongest is quantum simulation.  The idea of a quantum simulator has its origins in the early works of D. Deutsch and his Universal Quantum Turing Machine, which represents in quantum computability theory exactly what the Universal Turing Machine represents for classical computability. The idea of a universal quantum machine then led Lloyd in 1996 to prove that such a machine actually acts as a universal quantum simulator.

A quantum simulator requires by definition a discrete description of the phenomenon to be simulated, where discrete means a system made of disjointed components, e.g. qubits. The concept of discrete is different from the concept of discretized. In the latter we mean a system, at first continuous, resulting from a partition, seemingly arbitrary. If the phenomenon to be simulated is not discrete, the choice of finite element partitioning techniques is fundamental for the approximation to be sufficiently correct. Where "sufficiently" refers to the universal properties and symmetries of the original continuous system we are interested in. But these properties are generally not retained in the process of discretizing. 

Let us take two symmetries very common to continuous systems, translation invariance and isotropy. Lattice models replace the translation invariance with the invariance under translation of step equal to an integer multiple of the step size of the lattice.  Isotropy, which cannot be preserved in the discrete, is replaced by the invariance under the action of a finite group of rotations. Both these substitutions are well defined and in the limit in which the size of the lattice becomes various orders larger than the size of the elementary cell, they converge to the correct continuous symmetries.
%Using lattice models is especially relevant when investigating universal properties, that is properties common to physical systems of very different natures but sharing the same geometrical properties and symmetries. Lesne proved in 1998, using renormalization methods, that these are kept identical in continuous systems and in their discretized lattice versions, provided that the lattice is regular.
Evidently the debate on discretization procedures and the choice of the model cannot be limited to the symmetries and universal properties of the continuous system that is intended to be similar. Beyond computational techniques, there is also the choice of the scale of description and the nature of the phenomenon. 
 
In this thesis we will focus our attention on a class of quantum phenomena, described by transport equations which are linear partial differential equations (PDE), with constant or non-homogeneous coefficients. From a physical point of view this class describes a wide spectrum of transport phenomena, including the Dirac equation which is the one describing the motion of those particles called fermions, which constitute one of the two fundamental families into which all known particles are divided. Being able to simulate this equation therefore means being able to simulate the dynamics of one of the fundamental bricks of matter of which the universe is constituted. The need to simulate such phenomena comes from their algorithmic complexity in the multi-particle case, as already mentioned, but also from the fact that reproducing such dynamics in extreme regimes (e.g. in the presence of a strong magnetic field), is often impossible in the laboratory. It is therefore fundamental to simulate such regimes by means of controllable and easily accessible physical systems. 

There are essentially two methods to simulate a continuous quantum system on a lattice: one involves the series development of the continuous quantum operator, in terms of local units and its subsequent truncation to be implemented on a quantum computer, \emph{aka} trotterization. This kind of scheme by trotterizing the global operator in local unitaries is usually in continuous time and we referred to it as Hamiltonian simulation. Moreover, truncating an infinite series induces an error in the simulation that can be estimated and controlled.  An alternative way is the construction of a network of local quantum gates, uniformly distributed across space and time. This solution is part of the vast class of quantum cellular automata (QCA) and can be historically presented in both continuous and discrete time. Both models have their advantages and disadvantages: the first one, due to the truncation, can introduce the breaking of some symmetries and therefore the non-preservation of quantities fundamental for the phenomenon we want to describe, but it has undoubtedly the advantage of building an ad hoc quantum circuit that could result, after all, in a good approximation. The QCA are a universal quantum computation system: defining sufficiently general unitaries on the grid allows us to choose appropriate parameters to converge in the limit to the correct continuous evolution. Recently such models have been used to describe the propagation of free particles, e.g. fermions, in the presence of fields of various nature \cite{Dimolfetta2012ab, Debbasch2012aa, arnault2016quantum, di2016quantum, marquez2017fermion, arrighi2018dirac, hatifi2019quantum}, and also on arbitrary manifolds \cite{Dimolfetta2013aa, di2014quantum,DiMolfettarandom,marquez2017fermion, arrighi2019curved} and lastly interacting systems \cite{arrighi2020quantum, sellapillay2021staggered}. Such an idea is not new and goes back again to the brilliant Feynman, who first discretized the dynamics of fermions on a space-time checkboard \cite{Feynman_chessboard}.  

The growing enthusiasm towards these discrete models is not only due to the ability to simulate quantum dynamical systems that are difficult to recreate in a laboratory, but to the fact that they themselves can be considered toy models of the way transport can be defined on a fundamentally discrete space-time structure. The issue of combining a discrete quantum phenomenon with the concept of continuous space-time has been the most fascinating enigma in the community of physicists for decades. Computer scientists can shed new light on this great challenge, by bringing concepts such as universality, classification and complexity to the ever richer frontier zone, enabling new perspectives to the quest for answers and ultimately to the scientific progress. 

A further incentive to study such simulation models derives from a recent study, which for the first time has revealed how some natural mechanisms, such as the transport of a quantum particle, can naturally implement quantum algorithms, such as the Grover algorithm \cite{roget2020grover}. Simulating therefore a transport equation, might have as an additional result to suggest new ways in the short and medium term for the design of quantum algorithms. 

The interlacement between quantum simulation, physics and algorithmics is the framework within which this thesis is inscribed. The humble contribution that we want to make is in particular the consideration of a discrete model of lattice simulation for a wide family of dynamics described by transport equations, known as Quantum Walks (QWs), which may be also seen as the one-particle sector of a QCA in discrete time.  Systematically we will demonstrate how in the continuous limit, in other words when the size of the discretization step tends to zero, the QWs equations converge to a very general class of transport equations. In this regard, we will also discuss ways in which a new family of QWs, referred to as \emph{plastic}, might admit both a space and time limit, and a discrete space continuous time only limit, leaving the spatial grid to be discrete. This result is particularly important because it formally unifies for the very first time the Hamiltonian quantum simulators in continuous time we mentioned earlier, and those in discrete time, \textit{i.e.} the QCAs.  Finally we show that in order to simulate such transport equations, the grid is not an absolute necessity: instead they may be defined on an arbitrary triangulations of the Euclidean space, paving the way for new families of quantum simulators on arbitrary complex simplicia. 

\section*{The plan}

In chapter \ref{chap:quantumwalks} we will present QW on the grid in one and two spatial dimensions. Then in chapter \ref{chap:Continuous Limit1D} and chapter \ref{chap:Continuous Limit2D}, we will introduce the definition of Plasticity and give a general procedure to consider the limits in continuous time and discrete space and continuous space-time, respectively in 1D and 2D. In chapter \ref{chap:triangle}, we will introduce QWs on equilateral triangles and then generalize this result in chapter \ref{chap:manifold} by introducing QWs on arbitrary triangulations. In each of these chapters, we will systematically evaluate the continuous limit. In the last chapter \ref{chap:perspectives}, we illustrate some of the remaining steps to take, according to our judgement.  

\newpage
 %5 pages -- proofread
% 10 pages

\chapter{Elements of quantum theory}\label{chap:quantumtheory}

\toquote{So, what is quantum mechanics? [...] Basically, quantum mechanics is the operating system that other physical theories run on as application software (with the exception of general relativity, which hasn't yet been successfully ported to this particular OS). [...]But if quantum mechanics isn't physics in the usual sense – if it's not about matter, or energy, or waves, or particles – then what is it about? From my perspective, it's about information and probabilities and observables, and how they relate to each other.}{Scott Aaronson, Quantum Computing since Democritus (2013)}

\toabstract{In this chapter we will expose the fundamental principles of quantum mechanics alongside the mathematical formalism on which they are grounded, which is that of Hilbert spaces. In particular we will quickly present the necessary fundamentals of linear algebra, including the definitions of vector spaces, of scalar and tensorial products, the definition of linear operators. }

\clearpage

Before we introduce the Quantum Walks, we believe it is opportune to briefly present the principles of quantum mechanics, \textit{i.e.} the underlying mathematical structure of all quantum physical systems. It is not possible to provide in few pages a complete and exhaustive presentation, so we will limit ourselves to introduce only those ``game rules" useful to understand the content of this thesis, leaving it to the reader's curiosity to more complete reviews, such as \cite{NielsenChuang,cohen2006quantum}.

\section{Reminders of linear algebra }\label{sec:linalg}

\subsection{Vectors}\label{subsec:vectors}

\subsubsection*{Vector spaces, Inner product, Hilbert spaces}

\noindent A good understanding of quantum mechanics passes through a solid background of linear algebra fundamentals. In the following we will quickly recapitulate the elements necessary to introduce and understand the postulates of quantum theory. The basic objects of linear algebra are vector spaces. The vector space of interest for quantum theory is the space of all n-tuples of complex numbers $\mathbf{z} = (z_1, ..., z_n) \in \mathbb{C}^n$. The elements of a vector space are vectors and from now on we will indicate them following the notation most used by quantum computer scientists, that is, the \emph{Dirac notation}. According to this notation a vector named $z$ is called \emph{ket} and it is denoted by $\ket{z} \in \mathbb{C}^n$. The dual vector to $\ket{z}$ is called \emph{bra} and it is denoted by $\bra{z}$.  

\noindent The inner product for $\mathcal{V}$ is $(.,.):\mathcal{V}\times \mathcal{V}\lra \mathbb{C}$ if and only if it verifies the 
following set of axioms: 
\begin{align*}
\forall \ket{\psi} \forall \ket{\phi} 
\forall \ket{\chi}~(\ket{\psi},\ket{\phi}+\ket{\chi}) &= (\ket{\psi} , \ket{\phi}) + (\ket{\psi} , \ket{\chi})\\ 
\forall \ket{\psi} \forall \ket{\phi}~(\ket{\psi},\ket{\phi}) &= (\ket{\phi},\ket{\psi})^*\\
\forall \ket{\psi} ~(\ket{\psi},\ket{\psi}) &\geq 0 \textrm{ with equality iff } \ket{\psi}=\mathbf{0}. 
\end{align*}
In the Dirac notation the inner product $(\ket{\psi},\ket{\phi})$ is denoted simply by $\braket{\psi}{\phi}$.\\ 
Two vectors are \emph{orthogonal} if their inner product is zero. For example, if $\psi \equiv (0~1)$ and $\phi \equiv (1~0)$, then $\braket{\psi}{\phi} =0$, so they are orthogonal because their inner product is zero. Then, inner product induce the definition of the \emph{norm}: denoted by $||\ket{\psi}||$, is the non-negative real number $\sqrt{\braket{\psi}{\psi}}$.

\noindent Now let $S$ be a countable set, \textit{i.e.} $S$ is either finite or in bijection with $\mathbb{N}$. We denote by $\mathcal{V}_S$ the vector space generated by finite linear combinations of the vectors $(\ket{e})_{e\in S}$, and endowed with an inner product such that the $(\ket{e})_{e\in S}$ are orthonormal, \textit{i.e.} $\braket{e}{e'}=\delta_{{e}{e'}}$. Generally a vector space may have many different spanning sets $(\ket{e})_{e\in S}$. If such set of vectors are linearly independent then we will call this set a \emph{basis} of $\mathcal{V}$. Such basis always exists and the number of element in the basis is defined to be the dimension of $\mathcal{V}$. Finally we call a vector space equipped by an inner product an inner product space or \emph{Hilbert space}.\\

\subsubsection*{Tensor product}

\noindent The \emph{tensor product} is an operation of putting vector space together to form a larger vector spaces. Let us consider two Hilbert spaces $\mathcal{V}$ and $\mathcal{W}$ of dimension respectively $m$ and $n$. The tensor Hilbert space $\mathcal{V}\otimes\mathcal{W}$ is composed by all possible linear combinations of tensor product $\ket{v}\otimes\ket{w}$\footnote{We will often use the abbreviated notations $\ket{v}\ket{w}$ or $\ket{v w}$ for the tensor product $\ket{v}\otimes\ket{w}$.}, where $\ket{v}\in\mathcal{V} $ and $\ket{w}\in\mathcal{W} $. In particular if $\ket{e_v}$ and $\ket{e_w}$ are basis respectively of $\mathcal{V}$ and $\mathcal{W} $ then $\ket{e_v} \otimes \ket{e_w} $ is a basis of $\mathcal{V}\otimes\mathcal{W}$. The tensor product satisfies the following properties:
\begin{itemize}
\item For an arbitrary scalar $c$ and element $\ket{v}$ of $\mathcal{V}$ and  $\ket{w}$ of $\mathcal{W}$, 
\begin{equation}
c (\ket{v}\ket{w}) = (c \ket{v})\ket{w}) = \ket{v}(c\ket{w}). 
\end{equation}
\item For arbitrary $\ket{v_1}$ and $\ket{v_2}$ of $\mathcal{V}$ and  $\ket{w}$ of $\mathcal{W}$, 
\begin{equation}
(\ket{v_1}+\ket{v_2})\ket{w} = \ket{v_1}\ket{w}+\ket{v_2}\ket{w}.
\end{equation}
\item For arbitrary $\ket{w_1}$ and $\ket{w_2}$ of $\mathcal{W}$ and  $\ket{v}$ of $\mathcal{V}$,
\begin{equation}
\ket{v}(\ket{w_1}+\ket{w_2}) = \ket{v}\ket{w_1}+\ket{v}\ket{w_2}.
\end{equation}
\end{itemize}

\subsection{Operators}\label{subsec:operators}

Consider a Hilbert space $\mathcal{V}$ and an application $Q$ defined on $\mathcal{V}$ to values in $\mathcal{V}$, which also satisfies, for each pair of complex numbers $\lambda$, $\mu$, and ket $\ket{v}$ and $\ket{w}$ the property :
\begin{equation}
Q(\lambda \ket{v}+\mu \ket{w}) = \lambda Q\ket{v}+ \mu Q\ket{w}.
\end{equation} 
Then $Q$ is said to be a linear operator. In quantum theory we consider almost exclusively operators of the above mentioned form and for this reason in the following we will omit the adjective linear. 
Given two operators $Q$ and $R$ we can define the sum operator $R+Q$ and the operator $RQ$ in the following way:
\begin{equation}
\begin{split}
(R+Q) \ket{v} = R\ket{v}+Q\ket{v}.
(RQ) \ket{v} = R(Q\ket{v}) \equiv RQ \ket{v}.
\end{split}
\end{equation} 
whatever $\ket{v}$ is. Generally the product between two operators is not switchable and it is useful to consider this operator, called commutator:
\begin{equation}
[R,Q] = RQ - QR
\end{equation}
and the anti-commutator 
\begin{equation}
\{R,Q\} = RQ + QR.
\end{equation}
The reverse operator of a $Q$ operator is denoted $Q^{-1}$, and is such that
\begin{equation}
Q Q^{-1} = \mathds{1},
\end{equation}
where $\mathds{1}$ is the identity operator. \\
It is possible to define operators not only on the direct, `ket' vector space, but also on its dual space. Let $Q$ be a linear operator and consider 
\begin{equation}\label{eq:Q}
\bra{w}(Q\ket{v}).
\end{equation}
In general we can always associate to each linear operator $Q$ acting on kets an operator acting on bra. In particular if $Q$ can operate indifferently on ket or bra we will have:
\begin{equation}\label{eq:Q1}
\bra{w}(Q\ket{v}) = (\bra{w}Q)\ket{v} = \bra{w}Q\ket{v}.
\end{equation}
Now let us consider $\bra{w}Q$ and an other operator $Q^\dagger$. If 
\begin{equation}\label{eq:Q2}
(\ket{w},Q\ket{v}) = (Q^\dagger \ket{w},\ket{v})
\end{equation}
then $Q^\dagger$ is said to be the adjoint of $Q$. Let's remember some basic properties:
\begin{equation}
\begin{split}
(RQ)^\dagger = Q^\dagger R^\dagger\\
\bra{w}Q\ket{v}^* = \bra{v}Q^\dagger \ket{w}\\
(Q^\dagger)^\dagger = Q,
\end{split}
\end{equation}
where $<.>^*$ is the complex conjugate of $<.>$.\\
A definition of great importance in Hilbert spaces that has a central role in quantum mechanics is the following: an operator $Q$ that satisfies 
\begin{equation}\label{eq:herm}
Q=Q^\dagger
\end{equation}
is said to be self-adjoint or \emph{Hermitian}. Finally we will call \emph{unitary operator} an operator $U$ that satisfies 
\begin{equation}
UU^\dagger  = U^\dagger U = \mathds{1}.
\end{equation}
A unitary operator always has an inverse: $U^{-1} = U^\dagger$.

\subsubsection*{Eigenvectors and eigenvalues}

Consider $\lambda \in \mathbb{C}$. We call it an \emph{eigenvalue} of the operator $Q$ if there is a ket $\ket{\lambda}$ such that 
\begin{equation}
Q\ket{\lambda} = \lambda \ket{\lambda}.
\end{equation}
The ket $\ket{\lambda}$ is called the \emph{eigenvector} with eigenvalue $\lambda$. The set of eigenvalues of an operator is called the \emph{spectrum}. 

\subsubsection*{Outer product}

Given two vectors $\ket{v}$ and $\ket{w}$ the operator $P = \ket{v}\bra{w}$ is called outer product between $ \ket{v}$ and $\bra{w}$. It is defined as follows: For all $\ket{h}\in \mathcal{V}$ we have $P \ket{h} = \ket{v}\bra{w}\ket{h}$. Notice that, differently from the scalar product, the outer product is not a number but an operator. In particular, the operator $P =\ket{w}\bra{w}$ is called \emph{orthogonal projector} as it satisfies $P^2 = P$ and $P^\dagger = P$. The eigenvalues of projectors are $0$ or $1$, the eigenvector associated to $1$ is $\ket{w}$ and the one associated to $0$ is any other vector $\ket{\bar w}$ orthogonal to $\ket{w}$. Notice that projectors are more general than this, but the above one is the only one we will consider in this thesis. 

\section{Postulates of quantum theory }\label{sec:postulates}

Quantum mechanics is the theory of closed systems. By closed, we mean isolated \ie such that there is no interaction between the quantum particle and the outside. We will see that the evolution of closed quantum systems is linear and reversible. In fact whenever a closed system interacts with the outside world, by means of a large number of degrees of freedom, the system loses its `quantum behaviour'. This is why building a quantum computer is so challenging from a technological point of view. The first principle sets up the arena in which quantum mechanics takes place:

\begin{Post}[States]~\\
The state of any closed quantum system is fully 
described by a unit norm vector $\ket{\psi}$ in $\mathcal{H}$.
\end{Post}
Example. The simplest quantum mechanical system is the \emph{qubit}. Suppose that the Hilbert space $\mathcal{H}_2$ of the qubit $\ket{\Psi}$ is spanned by the orthogonal basis $\{\ket{0},\ket{1}\}$, then an arbitrary state in the state space is described by the generic superposition state $\Psi = a\ket{0} + b\ket{1}$ where the complex amplitudes $a$ and $b$ satisfy the normalisation condition $|a|^2+|b|^2=1$ such that $\langle\Psi |\Psi\rangle =1$. 

Since the qubit is isolated, it is somewhat natural that evolutions in time be norm-preserving over $\mathcal{H}$. This translates into the following postulate.
\begin{Post}[Evolutions]~\\
The evolution of a quantum system in discrete time steps is fully described by a unitary evolution $U$ over $\mathcal{H}$, \textit{i.e.} if a quantum system has state $\ket{\psi}$, then at the next time step it will have state $U\ket{\psi}$. 
\end{Post}
Let's look at some examples of unitary operator over a single qubit which are important for this thesis: (i) The \emph{Pauli Matrices}: 
\begin{equation}
\sigma_x = \begin{pmatrix} 0&1\\1&0\end{pmatrix},\hspace{1cm}\sigma_y = \begin{pmatrix} 0&-i\\i&0\end{pmatrix},\hspace{1cm}\sigma_z = \begin{pmatrix} 1&0\\0&-1\end{pmatrix}.
\end{equation}
Notice that $[\sigma_i,\sigma_j] =2 i \varepsilon^{ijk}\sigma_k$ where $\varepsilon^{ijk}$ is the Levi-Civita tensor (\ie $\varepsilon^{ijk} =1$ if the permutation of the indices is even, -1 if it is odd, 0 if the two indices $i$ and $j$ coincide). The $\sigma_x$ and $\sigma_z$ are also sometimes referred to as the \emph{bit flip} and the \emph{phase flip}. 
(ii) An other important unitary operator is the \emph{Hadamard gate}
\begin{equation}
H = \frac{1}{\sqrt{2}}\begin{pmatrix} 1&1\\1&-1\end{pmatrix}.
\end{equation}

Note that in the case when $\mathcal{H}$ is of finite dimension, any unitary operator $U$ can be approximated to an arbitrary precision by a composition (in time with the usual operator composition, and in space with the tensor product) of quantum gates chosen in a universal set, such as $\{\textsc{Cnot}, \textsc{H}$, $\textsc{S}$ and $\textsc{T}$\}\footnote{The phase gates $\textsc{S}$ and the $\textsc{T}$ read respectively
\begin{equation}
\textsc{S} = \begin{pmatrix} 1&0\\0&i\end{pmatrix},\hspace{1cm}\textsc{T} = \begin{pmatrix} 1&0\\0&e^{i \pi/4}\end{pmatrix}
\end{equation}}. See for instance \cite{BoykinGates,NielsenChuang}.\\

It is also useful to recall that this second principle can also be given in continuous time where the evolution of the quantum system is fully described by the Schr\"odinger equation 
\begin{equation}\label{eq:SEdef}
i \partial_t \ket{\psi} = \mathbb{H} \ket{\psi},
\end{equation}
where $\mathbb{H}$ is a fixed Hermitian operator known as the \emph{Hamiltonian} of closed system. Analogously to the discrete case, once we know the Hamiltonian of a system, we can determine its dynamics completely. Notice that if $\mathbb{H}$ is time independent then the solution of the trajectory of the system in the state space will be described by $\psi(t) = e^{- i \mathbb{H}(t-t_0))}\psi(t_0)$, where it is straightforward to show that the operator $U(t-t_0) = e^{- i \mathbb{H}(t-t_0))}$ is unitary. We can conclude that any unitary operator $U$ can be realised in the form $U = e^{i K}$ for some Hermitian operator $K$. In other words, there exists a one-to-one correspondence between the discrete-time description of the evolution and the continuous time one using the Hamiltonian. This consideration will be important all along this thesis. \\

As we already know the tensor product is the canonical way to take two Hilbert spaces and embed them in a bigger one. Hence it is natural that this should be the operation used in order to put two quantum systems together. 

\begin{Post}[Composite systems]\label{pos:composite}~\\
The state space of a composite physical system is the tensor
product of the state space of the component physical
systems.\\
Before the two systems have interacted in
any manner if we have $\ket{\psi}$ the state of
system $A$ and $\ket{\phi}$ the state of system $B$,
then the joint system is 
$\ket{\psi}\otimes\ket{\phi}$.\\
\end{Post}
Example. Consider the state $(\ket{0}+\ket{1})/\sqrt{2})\otimes \ket{0}$. By bilinearity this is equal to $(\ket{0}\otimes \ket{0}+\ket{1}\otimes \ket{0})/\sqrt{2})$. Now consider applying the unitary gate $\textsc{Cnot}$, which is defined to map $\ket{1}\otimes\ket{0}$ to $\ket{1}\otimes\ket{1}$, $\ket{1}\otimes\ket{1}$ to $\ket{1}\otimes\ket{0}$, and leave $\ket{0}\otimes\ket{0}$ and $\ket{1}\otimes\ket{1}$ unchanged. Then the state becomes  $(\ket{0}\otimes \ket{0}+\ket{1}\otimes \ket{1})/\sqrt{2})$. This is no longer a product state: it can no longer be written as $\ket{\psi}\otimes\ket{\phi}$ for any choice of $\ket{\psi}$ and $\ket{\phi}$. The two quantum systems, as they interacted through the $\textsc{Cnot}$ gate, have become `entangled'. This state lives in $\mathcal{H}^A\otimes \mathcal{H}^B$. 
\newpage
 %10 pages -- proofread
%11 pages

\chapter{Fundamentals of Quantum Walks}\label{chap:quantumwalks}

%\toquotebig{citation}{auth}

\toabstract{
We will consider discrete systems, living in discrete space and discrete time: We will focus on the dynamics of single particles, usually with some internal degrees of freedom. The key content of a discretization unit, or ``cell”, is whether the particle is in the cell or not, and what its internal state is. Moreover, as for any quantum system, these properties may be found in superposition. In a single time step the particle can only move a finite distance. These are the basic ingredients of a discrete-time quantum walk (DTQW). We will present this discrete model on the line and on the square lattice. 
}

\clearpage

Quantum walks are the quantum counterpart of the Classical random walks (CRWs), which are employed to model phenomena as chemical reactions \cite{gillespie1977exact, komkov1991random, van1992stochastic}, genetic sequence location \cite{van1992estimating, lange1991random, neigel1993application}, optimal search strategies \cite{lv2002search, bartumeus2005animal, oshanin2007intermittent}, diffusion and mobility in materials \cite{kirchheim1985modelling, galla1979two, tunaley1974asymptotic}, exchange rate forecasts in economical sciences \cite{macdonald1994monetary, campbell1997econometrics, kilian2003so} and information spreading in complex networks \cite{noh2004random, newman2005measure, rosvall2008maps}. Furthermore, they can successfully implement efficient algorithms, for example they can solve differential equations \cite{anderson1975random, higham2001algorithmic}, optimization \cite{bartumeus2005animal, bohachevsky1986generalized} and clustering problems \cite{rammal1983random, yen2005clustering}. Random walks spread to every domain of science for more than a century and are still an important source for researchers nowadays. \\

Even though in the second half of last century the interest in a quantum analogous to the classical stochastic process led to further investigation of quantum mechanics and quantum information, since the 1960s, numerous scientists \cite{schwinger61, iche78, grabert83, gardiner1985handbook, flindt2008counting, accardi2002quantum, james2008control, parthasarathy2012introduction} have extended the Brownian motion and stochastic calculus to particles which exhibited quantum effects. Schwinger was the first to demonstrate the importance of coherence effects in the evolution of a Brownian quantum particle. Just few years later, \cite{Fjeldso88} proposed the first discrete model with the intention of recovering a quantum version of the CRW. This first example represented a quantum planar rotor on a lattice whose dynamics can be approximated by a random walk. Another important milestone was the work of Gudder \cite{gudder90}, who studied systematically quantum Markov processes and established formally a connection with Feymann path integral formalism.  \\
 
Interestingly some years later, \cite{godoy92} and \cite{grossing1988quantum} had independently the same intuition of developing a quantum analogous to the CRW in discrete time and discrete space. Godoy and Fujita proposed a one-dimensional Markovian quantum walk displaying a diffusive behavior and \cite{grossing1988quantum} formalized the first model of unitary Quantum Walk (QW), whose dynamics was fully ballistic. Let us remark that, although the latter was defined as a QCA, this model is completely equivalent to the one proposed later by Aharonov \cite{Aharonov1993aa}. Indeed, QWs may be seen as the one particle sector of a QCA, extensively reviewed in \cite{arrighi2019overview}.\\

     Thanks to its features, especially the unitarity, QWs were immediately considered a new and efficient tool for solving, in a wider range of applications, technical problems in a more convenient way than classical random walks.

\section{Quantum Walks}

In this section we will introduce the simplest quantum walk on a Cartesian grid, i.e. over a monohedral tessellation of a Euclidean space by congruent squares (or unit cubes in higher dimensions), where the vertices are point on the integer lattice. We will start by defining a QW on an infinite discrete line and treat its higher dimensional extension later. 

\subsection{Definitions}\label{subsec:def1D}

We need few ingredients to define a QW in discrete time: the walker, its internal state and an evolution unitary operator for both walker and coin. 

\begin{Def}[The walker]\label{def:walker} ~ The walker is represented by a quantum state $\ket{\omega} $ living in a Hilbert space $\mathcal{H}_\mathbb{Z}$ of infinite but countable dimension. The computational basis of this space is $\ket{m} \in \mathbb{Z}$, $i.e.$, the position sites of the walker. The walker $\ket{\omega} $ will be then denoted by any superposition of the form:
\begin{equation}\label{eq:walker}
\ket{\omega}  = \sum_{m\in\mathbb{Z}} \alpha_m \ket{m}
\end{equation}
such that the normalisation condition $\sum_m |\alpha_m|^2 = 1$ holds. 
\end{Def}

The simplest but not trivial QW needs an internal state which we call coin state.

\begin{Def}[The coin]\label{def:coin} ~  The coin state $\ket{\kappa}$  is a quantum system, living in a $k-$dimensional Hilbert space $\mathcal{H}_\kappa$, spanned by the canonical basis $\{\ket{0},\ket{1},...,\ket{k-1}\}$. Thus $\ket{\kappa}$ may be written as 
\begin{equation}\label{eq:walker}
\ket{\kappa}  = \sum^{k-1}_s \beta_s \ket{s}
\end{equation}
where $\sum_s |\beta_s |^2 = 1$.
\end{Def}

In the following we will consider the coin state as belonging to a $2-$dimensional Hilbert space $\mathcal{H}_2$, as the most natural extension of the notion of the classical bit. Thus the overall state, representing the QW, lies in both Hilbert spaces and it is spanned by the coin state and the position state basis.

\begin{Def}[Quantum Walks over infinite line]\label{def:QW1D}~A Quantum Walk over the one dimensional grid in discrete time is represented by the state $\Psi$,  lying in the composite Hilbert space $\mathcal{H}_2\otimes \mathcal{H}_\mathbb{Z}$, as introduced in postulate \ref{pos:composite}. For the internal state we may choose some orthonormal basis $\{\ket{0}, \ket{1}\}$ and the overall state may be written 
\begin{equation}\label{eq:QW}
\Psi=\sum_m \psi^0_m \ket{0}\otimes\ket{m} + \psi^1_m \ket{1}\otimes\ket{m},
\end{equation}
where the complex probability amplitude $\psi^0$ and $\psi^1$ have to respect the following normalisation condition $\sum_m (|\psi^0_m|^2+ |\psi^1_m|^2) =1$.
\end{Def}

The temporal evolution of a quantum walk is driven by the composite action of two unitary operators, one acting in combined position-coin space and the other in the coin space. Similar to the classical random walk, we need one operator to move the walker on the line and one operator to play the same role as the coin toss. The latter is crucial in the dynamical features of the QW. Unlike the classic case, where such an operator is represented by a stochastic matrix, in the case of the QW evolution there is no room for randomness before measurement and it is represented by an unitary matrix which acts as an internal rotation in the internal state space.

\begin{Def}[Quantum Coin]\label{def:coin}~The most general operator acting on a two-dimensional state is an arbitrary element of the unitary group $U(2)$, depending on four real parameters $\alpha$, $\xi$, $\zeta$, and $\theta$ in the following way:
\begin{equation}\label{eq:rotation}
        \widehat C_{\theta, \alpha,\xi,\zeta}=e^{i\alpha}\begin{pmatrix}\cos\theta e^{i\xi} & \sin\theta e^{i \zeta}\\ -\sin\theta e^{-i \zeta} & \cos\theta e^{-i\xi}\end{pmatrix}
\end{equation}
\end{Def}

Taking advantage again of the analogy with a classical random walk, the shift operator, once the coin is flipped, moves the walker one step in a precise direction, depending on the state of the coin. While, in the classic case, the walker moves left or right, depending on whether the result is heads or tails, in the case of the unitary evolution of a QW, the walker will move in superposition of states:

\begin{Def}[Shift Operator]\label{def:shift}~
A suitable conditioned shift operator for the QW has the form:
\begin{equation}
\widehat S = \ket{0}\bra{0}\otimes \sum_{m}\ket{m+1}\bra{m} + \ket{1}\bra{1}\otimes \sum_{m}\ket{m-1}\bra{m}.
\end{equation}
\end{Def}

At last we can introduce the unitary evolution operator acting on the Hilbert space $\mathcal{H}_2\otimes \mathcal{H}_\mathbb{Z}$.

\begin{Def}[Evolution operator]\label{def:evo1D}~
Let us introduce $j \in \mathbb{N}$ to label instants and let the QW prepared for the initial time $j'$ as a general product state \ref{eq:QW}
then

\begin{equation}\label{eq:evo1D}
\Psi_{j'+1} = \widehat W \Psi_{j'}
\end{equation}
where 

\begin{equation}
\widehat W = \widehat S (\widehat C_{\theta, \alpha,\xi,\zeta} \otimes \mathds{1}_\mathbb{Z})
\end{equation}
is the evolution unitary operator. 
\end{Def}

Notice that, due to the unitarity of $W$, the $\rho_{j'} \equiv |\Psi_{j'}|^2$ is a conserved real number all along the evolution and it is interpreted as the probability density of measuring the walker's being detected at any point of the space. It is usually normalised to one.  

As an example, we will study the Hadamard Walk, the most common one-dimensional QW. 

\begin{Ex}[Hadamard Walk]\label{ex:Had}
In the Hadamard Walk, the coin is the Hadamard operator, which coincides with $\widehat H = \widehat C_{\frac{\pi}{4}, \frac{\pi}{2},\frac{-\pi}{2},\frac{-\pi}{2}}$:
\begin{equation}
\widehat H = \frac{1}{\sqrt{2}}\begin{pmatrix} 1 & 1 \\ 1 & -1\end{pmatrix} 
\end{equation} 
One time step of the Hadamard Walk consists to apply to the initial state $\Psi_{j'}$ the Hadamard coin and then the shift operator \ref{def:shift}. Given the following generic initial state:
\begin{equation}\label{eq:initialgeneric}
\Psi_{j'} = \sum^1_{s =0} \sum_{m=\infty}^\infty \psi^s_{m,j'} \ket{s}\ket{m}
\end{equation}
the state at instant $j'+1$ reads:
\begin{equation}
\begin{split}
\Psi_{j'+1} =  \sum_{m=\infty}^\infty\widehat S ( \psi^0_{m,j'} \widehat H\ket{0}\ket{m} +\psi^1_{m,j'} \widehat H \ket{1}\ket{m}) = \\
=  \sum_{m=\infty}^\infty\left( \frac{\psi^0_{m,j'}+\psi^1_{m,j'} }{\sqrt{2}}\ket{0}\ket{m+1} + \frac{\psi^0_{m,j'}-\psi^1_{m,j'} }{\sqrt{2}}\ket{1}\ket{m-1}\right)
\end{split}
\end{equation}
From the above equation, once the final state is written as linear combination in the computational basis, we can extract the recursive relations for the coefficients, i.e., the walker probability amplitudes:
\begin{equation}\label{eq:recursive}
\begin{split}
\psi^0_{m,j'+1} =  \frac{\psi^0_{m+1,j'}+\psi^1_{m+1,j'} }{\sqrt{2}}\\
\psi^1_{m,j'+1} =  \frac{\psi^0_{m-1,j'}-\psi^1_{m-1,j'} }{\sqrt{2}}\\
\end{split}
\end{equation}
For a given initial condition, the above equation can be solved numerically, generating the probability distribution and computing all the statistical property of the walker as the expected distance from the origin, i.e. the standard deviation, and other statistical momenta. 
%Let us choose a fair initial state $\Psi_0 = \ket{0}\ket{0}$ localised at the origin: 
%\begin{equation}
%\Psi_1 = \widehat W \ket{0}\ket{0} =\frac{1}{\sqrt{2}}( \ket{0}\ket{1} + \ket{1}\ket{-1})
%\end{equation}
%One time iteration of the Hadamard operator produces the superposition of the system both in position $m=1$ and $m=-1$. Notice that the above final state is symmetric and this fact depends on the choice of the coin and on the initial state. Starting with $\Psi_0 = \ket{1}\ket{0}$ would have introduce a minus sign in the final state producing an asymmetric state. 
\end{Ex}

\subsection{Dynamics}\label{subsec:prop}

There are several way to study analytically Eq. \ref{eq:evo1D} and extract information. Historically two approaches have been used: (i) the Schr{\"o}dinger approach and (ii) the combinatorial approach. The first one is the most used: starting from an arbitrary state of the QW, a Fourier transform of the operator leads to closed form of the coin amplitudes, guaranteeing a straightforward way to calculate the dynamical and statistical properties of the probability distribution. In the second one, given one point in the spacetime diagram $\ket{\bar m}$, we sum up all possible discrete paths ending up in $\ket{\bar m}$ starting from the initial condition. The latter approach, although is less used and mathematically more complicate, is reminiscent of the standard path integrals widely used in physics to study quantum particle behavior. All along this manuscript we will prioritize the first method, which will be carefully introduced in the next section. Since for the most general case \ref{def:evo1D} the evolution equations are very complex, we will focus on the Hadamard Walk on the line. 
%If our aim is to solve Eq. \ref{eq:recursive} analytically, then we have to address the problem in a different way. There exists a special basis, called Fourier basis, which diagonalise the shift operator diagonal, making possible a complete analytical study. 

\begin{Def}[Fourier transform]\label{def:fourier}~
The Fourier transform of a discrete function $g: \mathbb{Z} \rightarrow \mathbb{C} $ is a continuous function $\tilde g: [-\pi,\pi] \rightarrow \mathbb{C}$ defined by
\begin{equation}
\tilde g(k) = \sum_{m=-\infty}^{\infty} e^{-i k m} g_m .
\end{equation} 
The inverse transform is given by 
\begin{equation}
g_m = \int_{-\pi}^{\pi} \frac{dk}{2 \pi} \tilde g(k) .
\end{equation} 
\end{Def}

Using \ref{def:fourier} we can transform the computational basis of the walker Hilbert space $\mathcal{H}_\mathbb{Z}$, introducing the Fourier basis:

\begin{Def}[Fourier basis]\label{def:fourierbasis}~
Let us introduce the following vectors: $\{\ket{l_k}: -\pi \leq k \leq \pi\}$ and such that. 
\begin{equation}
\ket{l_k} = \sum_{m=-\infty}^\infty e^{i k m} \ket{m}
%Notice that $\ket{l_k}$ has infinity norm. 
\end{equation}
that we call Fourier basis. 
\end{Def}

The above definition \ref{def:fourierbasis} is used to decompose $\Psi_{j'}$ in the Fourier basis, as follows:

\begin{equation}
\Psi_{j'} = \int_{-\pi}^{\pi} \frac{dk}{2 \pi} \sum_{s=0}^{1} \tilde \psi^s_{j'}(k) \ket{s}\ket{l_k},
\end{equation}
where $\tilde \psi^i_j(k) $ are the coefficients of the vector $\Psi_j$ in the new basis. Now we have all the ingredients to prove the following Lemma:

%%%%From here I will use the letter a for the spin state, need to change above

\begin{Lem}{}\label{lem:FourierS}
The action of the shift operator $\widehat S$ on the overall state of the QW in the Fourier basis $\{\ket{l_k}: -\pi \leq k \leq \pi\}$ is the following:
\begin{equation}
\widehat S_k \ket{s}\ket{l_k} = e^{-(-1)^{s} i k} \ket{s}\ket{l_k}
\end{equation}
where $\widehat S_k$ is the shift operator.
\end{Lem}
\textbf{Proof.} 
\begin{equation}
\widehat S_k \ket{s}\ket{l_k} = \sum_{m=-\infty}^{\infty} e^{i k m}\widehat S \ket{s}\ket{m} = \sum_{m=-\infty}^{\infty} e^{i k m}\ket{s}\ket{m + (-1)^{s}}.
\end{equation}
Relabeling the discrete variable $m$ as $m'=m+(-1)^{s}$, we recover:
\begin{equation}
\widehat S_k \ket{s}\ket{l_k} = \sum_{m=-\infty}^{\infty} e^{i k (m'-(-1)^{s})} \ket{s}\ket{m'} = e^{-(-1)^{s} i k} \ket{s}\ket{l_k}.
\end{equation}
\begin{Rk}
The above Lemma leads to an explicit expression for the operator $\widehat S_k$. In fact, from the last equation and expanding on the computational basis, we have: 
\begin{equation}
\widehat S_k   = e^{-(-1)^{s} i k}=( e^{- i k}\ket{0}\bra{0} + e^{ i k}\ket{1}\bra{1} ),
\end{equation}
where $a = 0,1$. Thus,
\begin{equation}
\begin{split}\label{eq:sigmazDEF}
%\left(\cos(k) - i \sin(k)\right)\ket{0}\bra{0} +\left(\cos(k) + i \sin(k)\right)\ket{1}\bra{1} ) = \\
\widehat S_k  &= \cos(k)(\ket{0}\bra{0} +\ket{1}\bra{1})- i  \sin(k)(\ket{0}\bra{0} -\ket{1}\bra{1}) = \\
 &\cos(k)\mathds{1} - i  \sin(k)\sigma_z = e^{- i \sigma_z k},
\end{split}
\end{equation}
where in the last equation we used the Euler's formula and the definition of the third Pauli matrix $\sigma_z =(\ket{0}\bra{0} -\ket{1}\bra{1})$ decomposed in the computational basis $\{\ket{0},\ket{1}\}$.
\end{Rk}

The above Lemma shows that the shift operator acts in the Fourier basis as a phase, which also means that $\ket{s}\ket{l_k}$ is the eigenvector of $\widehat S$ associated to the eigenvalue $e^{-(-1)^{s} i k}$. \\

Now let us consider the overall operator $\widehat W_k$ in the Fourier basis. Using Eq. \ref{eq:sigmazDEF}, we obtain: 
%The last step to complete the diagonalise is to consider the Hadamard coin. Because all coefficients are constant, diagonalise $H$ is straightforward and we will omit here the technical details, even though it is still useful show the matrix representation of the evolution operator $W_k$ in Fourier basis:
\begin{equation}\label{eq:opFourier}
\widehat W_k = e^{- i \sigma _z k} \widehat H = \frac{1}{\sqrt{2}}\begin{pmatrix} e^{-i k} & e^{-i k} \\ e^{i k} & -e^{i k}  \end{pmatrix},
\end{equation}
The characteristic polynomial of $W_k$ is
\begin{equation}
\Pi(\lambda) = \lambda^2 + i \sqrt{2} \lambda \sin k - 1 
\end{equation}
and the eigenvalues are
\begin{equation}
\begin{split}
\alpha_k = e^{-i \omega_k},\\
\beta_k = e^{i (\pi + \omega_k)},
\end{split}
\end{equation}
where $\omega_k \in [\frac{\pi}{2},\frac{\pi}{2}]$ are the angles satisfying the equation
\begin{equation} \label{eq:omega}
\sin \omega_k = \frac{1}{\sqrt{2}} \sin k.
\end{equation}
The eigenvectors read 
\begin{equation}\label{eq:evec}
\begin{split}
\ket{\alpha_k} = \frac{1}{c^{-}} \begin{pmatrix}e^{-ik} \\ \sqrt{2} e^{-i\omega k} - e^{-ik} \end{pmatrix},\\
\ket{\beta_k} = \frac{1}{c^{+}} \begin{pmatrix}e^{-ik} \\ -\sqrt{2} e^{i\omega k} - e^{-ik} \end{pmatrix},
\end{split}
\end{equation}
where $c^{\pm} = 2 ( 1+ \cos^2 k) \pm 2 \cos k \sqrt{1 + \cos^2k}$. \\

Finally the evolution operator is decomposed as follows:
\begin{equation}
\widehat W = \int_{-\pi}^{\pi} \frac{dk}{2 \pi} \left(e^{-i \omega_k} \ket{\alpha_k,l_k}\bra{\alpha_k,l_k} + e^{i (\omega_k + \pi)}\ket{\beta_k,l_k}\bra{\beta_k,l_k} \right)
\end{equation}
and after $\tau$ time steps it reads simply as: 
\begin{equation}\label{eq:opFourier}
\widehat W^\tau= \int_{-\pi}^{\pi} \frac{dk}{2 \pi} \left(e^{-i \tau \omega_k} \ket{\alpha_k,l_k}\bra{\alpha_k,l_k} + e^{i \tau (\omega_k + \pi)} \ket{\beta_k,l_k}\bra{\beta_k,l_k} \right)
\end{equation}
Observe that the operator is diagonal in the Fourier basis which make simple the time integration of the QW equations. 

As conclusion of this section we can prove the following theorem: 

\begin{Th}\label{th:anasol}
Let us consider a Hadamard walker located at the origin $m = 0$ and the coin state with coin state $\ket{0}$. After $\tau$ time steps, the probability amplitudes of the walker read:
\begin{equation}
\begin{split}
\psi^0_{\tau,m} &= \int^\pi_{-\pi} \frac{dk}{2 \pi} \left( 1 + \frac{\cos k}{\sqrt{1 +\cos^2 k}} \right) e^{- i (\omega_k \tau - k m)},\\
\psi^1_{\tau,m} &= \int^\pi_{-\pi} \frac{dk}{2 \pi} \left( \frac{e^{i k}}{\sqrt{1 +\cos^2 k}} \right) e^{- i (\omega_k \tau - k m)},
\end{split}
\end{equation} 
where $\omega_k$ are defined in $\ref{eq:omega}$.
\end{Th}
\textbf{Proof.} 
The diagonal expression of \ref{eq:opFourier} is used to calculate the QW vector state at a generic time $\tau$:
\begin{equation}
\Psi_\tau = \widehat W^\tau \Psi_0 =  \int_{-\pi}^{\pi} \frac{dk}{2 \pi} \left(e^{-i \tau \omega_k} \ket{\alpha_k,l_k}\bra{\alpha_k,l_k}\ket{0,0}+ e^{i \tau (\omega_k + \pi)} \ket{\beta_k,l_k}\bra{\beta_k,l_k}\ket{0,0} \right) 
\end{equation}
where $\Psi_0 = \ket{0,0}$.  From Eq. \ref{eq:evec} we can compute
\begin{equation}
\begin{split}
\bra{\alpha_k,l_k}\ket{0,0} = \frac{e^{ik}}{\sqrt{c^-}},\\
\bra{\beta_k,l_k}\ket{0,0} = \frac{e^{ik}}{\sqrt{c^+}}.
\end{split}
\end{equation}
Thus, 
\begin{equation}
\Psi_\tau = \int_{-\pi}^{\pi} \frac{dk}{2 \pi} \left( \frac{e^{-i (\tau \omega_k - k)}}{\sqrt{c^-}} \ket{\alpha_k, l_k}+ \frac{e^{i \tau (\omega_k + \pi) + i k} }{\sqrt{c^+}}\ket{\beta_k,l_k} \right) \ket{l_k}.
\end{equation}
Let us now write the eigenvector in the computational basis $\ket{0},\ket{1}$:
\begin{equation}
\Psi_\tau = \int_{-\pi}^{\pi} \frac{dk}{2 \pi} \left(\frac{e^{-i (\tau \omega_k - k)}}{c^-} \begin{pmatrix}e^{-i k}\\ \sqrt{2} e^{-i \omega_k - e^{-ik}}\end{pmatrix}+ \frac{e^{i \tau (\omega_k + \pi) + i k} }{c^+}\begin{pmatrix}e^{-i k}\\ -\sqrt{2} e^{i \omega_k - e^{-ik}}\end{pmatrix}\right) \ket{l_k}
\end{equation}
and we can finally extract from the above equation the coefficients $\tilde \psi^{a}_\tau(k)$:
\begin{equation}
\begin{split}
\tilde \psi_\tau^0(k) &= \frac{e^{-i \omega_k t}}{c^-} + \frac{e^{i \tau (\omega_k + \pi) + i k} }{c^+},\\
\tilde \psi_\tau^1(k) &=\frac{e^{- i \omega_k \tau + i k} (\sqrt{2} e^{-i \omega_k} - e^{- i k})}{c^-} - \frac{e^{i(\pi + \omega_k)\tau + i k} (\sqrt{2} e^{i \omega_k} + e^{- i k})}{c^+}.
\end{split}
\end{equation}
Making explicit $c^\pm$ and after simplification we obtain:
\begin{equation}\label{eq:fourierCoef}
\begin{split}
\tilde \psi_\tau^0(k) &= \frac{1}{2} \left( 1 + \frac{\cos k}{\sqrt{1+\cos^2k}}\right)e^{- i \omega_k \tau} + \frac{(-1)^\tau}{2} \left(1- \frac{\cos k}{\sqrt{1+\cos^2k}} \right) e^{i \omega_k \tau}, \\
\tilde \psi_\tau^1(k) &= \frac{e^{ik}}{2 \sqrt{1+\cos^2 k}}(e^{-i\omega_k \tau} - (-1)^{\tau}e^{i \omega_k \tau}) \end{split},
\end{equation}
and finally the inverse Fourier transform of the above equation proves the theorem. \\

\begin{Rk}
It is important to notice that when we Fourier transform \ref{eq:fourierCoef} we recover two terms and depending on the value of $\tau$ these terms sum up or cancel. In particular when $m+\tau$ is odd the coefficient are always vanishing. This fact suggests that we can decompose the lattice in two sub-lattices, one in which $\tau$ and $m$ are both odd and one in which $\tau$ and $m$ are both even. This property is very general and does not depend on the choice of the initial condition or of the quantum coin. \end{Rk}

\section{Higher dimensional spaces}\label{subsec:defhigher}

%Higher dimensional lattices that lie in a plane can be built in a number of ways
The Eq. \ref{eq:evo1D} may be extended to any $d$-dimensional regular grid, e.g. square in 2D and cubes in 3D. 
%any monohedral tessellation of n-dimensional Euclidean space by regular and uniform polytopes, e.g. square or triangles in 2D and cubes or tetrahedra in 3D. 
The simplest way to generalize the previous section is to introduce a Hilbert space  for the position of the walker $\mathcal{H}_{\mathbb{Z}^d}$, where $d$ is the connectivity of the vertices, and  $\mathcal{H}_{d}$ for the coin state. The coin operator is then a $d$-dimensional matrix and the shift operator has to account a displacement towards the $i^{th}$ neighbor among the $d$ ones starting from a vertex. For instance, on a square lattice each vertex has four links and four neighbors, then the coin state encodes the probability amplitude to go towards each of the four possible directions. In a $d$-dimensional grid the most common one for the quantum coin is the Grover operator, introduced by Moore and Russell \cite{moore2002quantum}: 
\begin{Def}[Grover coin]
A $d$-dimensional Grover coin has elements $[C_d^{G} ]_{i,j} = \frac{2}{d} - \delta_{i,j}$, i.e.:
\begin{equation}
C_d^{G} = \begin{pmatrix} 2/d -1 & 2/d &\cdot & 2/d \\ 2/d  & 2/d-1 &\cdot & 2/d \\ .  & . & . & . \\ .  & . & . & . \\ .  & . & . & . \\   2/d  & 2/d &\cdot & 2/d-1\end{pmatrix}
\end{equation}
\end{Def}
Notice that, except for $d=4$, all Grover coins are biased since the diagonal entry is weighted differently from the off-diagonal ones. This particular coin has raised the interest of the community because it can be used to implement the two-dimensional Grover search algorithm \cite{shenvi2003quantum}. 

As an example, let us consider the square grid:  at each point of this infinite two-dimensional lattice, the walker has four possibilities to move ($d=4$), which translates in four possible coin states. Such a QW lies in a composite Hilbert space $\mathcal{H}_{square} = \mathcal{H}_{4} \otimes \mathcal{H}_{Z^2}$ where 
$ \mathcal{H}_{4}$ is spanned by the orthonormal basis $\{\ket{0},\ket{1}, \ket{2}, \ket{3}\}$. The shift operator is defined as follows:
\begin{equation}
\widehat S = \sum_{s=0}^{1}\sum_{m,n}\left( \ket{s}\bra{s} \ket{m+(-1)^{s},n}\bra{m,n}\right) +  \sum_{s=2}^{3}\sum_{m,n}\left( \ket{s}\bra{s} \ket{m,n + (-1)^{s}}\bra{m,n} \right)
\end{equation}
and for the coin we can choose the Grover coin:
\begin{equation}\label{eq:grover4}
C_4^{G} = \frac{1}{2}\begin{pmatrix} -1 & 1 & 1 & 1 \\ 1 & -1 & 1 & 1 \\ 1 & 1 & -1 & 1\\ 1 & 1 & 1 & -1 \end{pmatrix}.
\end{equation}
Although the above walk completely describes the walker spread on the plane, and it seems to be the most natural and straightforward generalization of \ref{eq:evo1D} in a spatial dimension higher then one, the use of a four dimensional state or two distinct qubits is really costly and nowdays very few feasible implementations have been proposed so far. To reduce the technological challenge, an alternative definition of 2D-dimensional QW has been introduced, which has been proven equivalent to the above one in the propagative regime \cite{di2011mimicking}. This new family of QW, that we call alternate quantum walk (AQW) is a composition of two one-dimensional QW, each moving in the orthogonal $u_x$ and $u_y$ directions. The quantum coin is a single qubit, then is spanned by $\{\ket{0}, \ket{1}\}$
The time evolution of the QW will be then driven by the application of the coin operator, e.g., the Hadamard coin, and the shift operator along each basis vector of the grid ${u_i}$:
\begin{equation}
S_{u_x} = \sum_{m,n} \left( \ket{m-1,n,0}\bra{m,n,0} + \ket{m+1,n,1}\bra{m,n,1} \right)
\end{equation}
and 
\begin{equation}
S_{u_y} = \sum_{m,n} \left( \ket{m,n-1,0}\bra{m,n,0} + \ket{m,n+1,1}\bra{m,n,1} \right)
\end{equation}
and in conclusion of this chapter, we give finally the following definition:

\begin{Def}[Alternate Quantum Walk]\label{def:alternateQW}~
The Alternate Quantum Walk is defined by the following unitary evolution 
\begin{equation}\label{eq:evo2D}
\Psi_{j'+1} = \widehat W \Psi_{j'}
\end{equation}
and 
\begin{equation}\label{eq:alternate2D}
\widehat W =  \left(\widehat S_{u_x} (\widehat H \otimes \mathds{1}_{\mathbb{Z}}) \widehat S_{u_y} (\widehat H \otimes \mathds{1}_{\mathbb{Z}})\right)
\end{equation}
\end{Def}

Remark that using a single qubit system for the walker means that the \textit{alternating} procedure can work only when $d$ is even. For instance in the case of a hexagonal lattice, where $d=3$, there not exist any splitting techniques for reducing a $3$-dimensional system to a sequence of single qubit ($2$-dimensional) quantum walk.

\medskip
\medskip
\medskip
\medskip
\medskip
\medskip
\medskip
\medskip

\noindent We have seen how QWs may be defined on a general $d$-dimensional regular grid. Plus we have given an overview of the most common techniques employed to study their dynamics. In the next Chapter we will introduce a new family of QWs, and we will introduce the main research problem of this manuscript, \ie the role played by QWs in simulating hyperbolic partial differential equations. 

\newpage
 %11 pages -- proofread
%14 pages

\chapter{Continuous limits and plasticity in 1D}\label{chap:Continuous Limit1D}

%\toquotebig{CITE}

\toabstract{
Discrete time implies that local dynamics will be generated by unitary discrete operations (or ”gates”). The non-trivial topology of the unitary group then leads to new dynamical possibilities, which have hardly been explored. A key question in all of these approaches is that of the continuum limit: For a quantum simulator, a limit statement is actually the success criterion, and for fundamental physics considerations such a statement plays the role of the correspondence principle. We will explore all families of Quantum Walks admitting both a continuous time discrete space and a continuous spacetime limit. These will be referred to as Plastic Quantum Walks on the line.}

\clearpage

The idea of simulating quantum systems with quantum computers was notably pointed out by Feynman~\cite{Feynman1965aa}, confronted with the insufficiency and intractability of classical computers' abilities to simulate quantum features. Several quantum simulation schemes have been sparked over the last few decades ~\cite{georgescu2014quantum} and most of them can be divided into two different approaches: some of the methods being used for simulating quantum systems implemented over \textit{discrete space} and \textit{continuous time} lattices consist of constructing a Hamiltonian, which is historically the continuous time operator governing the time evolution of the system, which imitates a physical system, or trotterizing a constructed Hamiltonian to obtain unitaries~\cite{jordan2012quantum,StrauchCTQW}. Problems with these approaches are discussed in Ref.~\cite{molfetta2019quantum} and include the breaking of Lorentz covariance as well as issues arising when recovering a bounded speed of light. The other approach concerns \textit{discrete spacetime models}, which do not share the difficulties of their discrete space continuous time counterparts, such as the quantum circuit model and the quantum cellular automata or their one-particle sector, namely the discrete time quantum walk (DTQW). In particular this last quantum scheme have been proved a powerful tool to simulate a large spectra of quantum natural phenomena which are usually described by partial differential equations (PDE): indeed the continuous spacetime limit of various DTQWs defined on the regular lattice in arbitrary dimensions has been proved to be equivalent to a wide range of hyperbolic PDE (HPDE) describing in general quantum transport, eventually coupled with abelian \cite{di2012discrete,MolfettaDebbasch2014Curved,Arnault_2016} and non-abelian gauge field \cite{di2016quantum,arnault2016quantum,ARNAULT2016179} on curved spacetime \cite{di2013quantum,ArrighiGRDirac3D,succiQWBoltzmann,arrighi2019curved}. 

The connection between continuous time and discrete space scheme with quantum walks has been investigated by \cite{StrauchCTQW, childs2010relationship, shikano2013discrete, dheeraj2015continuous}. Although several techniques have been employed to study the connections between them, these two models have been apart for a long time. Recently a quantum simulation scheme known as a Plastic Quantum Walk (PQW) have been developed ~\cite{molfetta2019quantum} which supports both a continuous spacetime limit and a continuous time-discrete space limit.  Plasticity is the definitive element that makes the continuous limit process less rigid and therefore plastic. While remaining on the space-time grid, a PQW adapts anisotropically, modifying the speed at which space and discrete time converge to the continuous. This new quantum scheme not only encapsulates both historical approaches to simulate a quantum system, but unveils a previously inconsiderate property of the space-time grid. 

In this chapter, we will explore plasticity in two space-time dimensions and we leave a complete treatment of the three space-time dimensional case to the next chapter. 

\section{Plastic Quantum Walk}\label{eq:CL1D}

A large class of QW admits the continuous limit in space-time and there are few examples that admit the limit in time, leaving the space discrete. Here, we want to demonstrate that by introducing anistropic scaling on the space-time grid, we can find QWs that admit both limits. Such QWs will be called plastics. 

\begin{Def}[Plastic Quantum Walk]
A Plastic Quantum Walk (PQW) is a QW which admits both a continuous limit in time and a continuous limit in spacetime depending on some parameters.
\end{Def}

We will begin to study the simplest but not trivial case of Plastic QW, and in the results we will define and discern strong and weak forms of plasticity and then end up being interested, in the rest of the text, only in the former. 

\subsection{The infinite line}

We consider a QW in one spatial dimension as defined in \ref{def:QW1D}. To investigate the continuous limit, we first introduce a time discretization step $\Delta_t$ and a space discretization step $\Delta$. We then introduce, for any discrete function $\psi$ appearing in the Definition ~\ref{def:evo1D}, a complex function $\bar \psi \in \mathbb{C}$ over the spacetime positions $\mathbb{R}^+ \times \mathbb{R}$, such that $\psi_{j,m}=\bar \psi(t_j,x_m)$, with $t_j=j \Delta_t$ and $x_m = m \Delta$. In other terms, we assume that the discrete function $\psi$ samples exactly the continuous complex function $\bar \psi(t,x)$.\\

The Eq. ~\ref{eq:evo1D} after a number of time steps $\tau$ reads: 
\begin{equation}\label{eq:1stepCont}
\bar{\Psi}(t_j+\tau \Delta_t) =  \widehat W^\tau \bar{\Psi}(t_j),
\end{equation}
where 
\begin{equation}\label{eq:newcoin}
\widehat W^\tau=(\widehat S_{u_x} (\widehat C_{\theta, \alpha,\xi,\zeta} \otimes \mathds{1}_\mathbb{Z}))^\tau.
\end{equation}

Now let us drop the bars to lighten the notation. We suppose that all functions are at least $C^2$, \ie twice differentiable. The spacetime continuum limit, when it exists, coincides with the coupled differential equations obtained from Eq. \ref{eq:1stepCont} by letting both $\Delta_t$ and $\Delta$ go isotropically to zero, as for example in \cite{di2012discrete}.

Before continuing, we will remember below the statement of Taylor's theorem that in mathematical analysis, is a theorem that provides a sequence of approximations of a function, assumed differentiable around a given point by Taylor's polynomials, whose coefficients depend only on the derivatives of the function at the point \cite{taylor1717methodus}. 

\begin{Th}[Taylor's theorem]
Let us consider an interval $(a,b)\subset \mathbb{R}$ and a point $x_0 \in (a,b)$. Be $f:(a,b)\to \mathbb{R}$ derivable $n-1$ times in the interval $(a,b)$, with $n\geq 1$ et let us assume that the $n^{th}-derivative$ $\partial^{(n)}f$ be continuous in $x_0$. Be the Taylor's polynomial of rank $n$ reads: 
\begin{equation}\label{eq:polyTaylor}
T_n(f,x) = f(x_0) + (x-x_0) \partial^{(1)}f + \frac{(x-x_0)^2}{2!} \partial^{(2)}f + ...= \sum^{n}_{i=0}\frac{(x-x_0)^i}{i!} \partial^{(i)}f.
\end{equation}
The approximated function of order $n$ of the function f is given by:
\begin{equation}
f(x) = T_n(f,x) + R_n(x),
\end{equation}
where is defined such that 
\begin{equation}
R_n(x) = \lim_{x\to x_0} \frac{R_n(x)}{(x-x_0)^n}=0
\end{equation}
or equivalently $R_n(x) = o(x-x_0)^n$, where we used the standard little $o-$notation.
\end{Th}

The Taylor's theorem will be used all along of the thesis to certify that the QW converge correctly to the continuous equations we wish to simulate. \\

Now coming back to plasticity, we will prove in the following that there is no room for plasticity in one spatial dimension if we compute the limit at each time iteration of the QW, for $\tau = 1$ in Eq. \ref{eq:opFourier}.

\begin{Th}[No plasticity for $\tau=1$]
Let us parametrize the time and space steps with a positive real number $\varepsilon \in \mathbb{R}^+$ and let consider the following jets:
\begin{equation}\label{eq:scalings}\begin{split}
\theta = \theta^{(0)}   + \varepsilon^b \theta^{(1)}\\
\alpha = \alpha^{(0)}   + \varepsilon^b  \alpha^{(1)}\\
\xi = \xi^{(0)}   + \varepsilon^b  \xi^{(1)}\\
\zeta = \zeta^{(0)}   + \varepsilon^b  \zeta^{(1)}\\
\end{split}
\end{equation} 
so that 
\begin{equation}\label{eq:Cjet}
\widehat C_{\alpha, \theta,\xi, \zeta} = \widehat C^{(0)}+ \varepsilon^b \widehat C^{(1)}
\end{equation}
where the exponent $b$ is a real positive and smaller then 1. Let $\Delta_t = \varepsilon$ and $\Delta = \varepsilon^{1-a}$, where $a$ is a real positive, such that $0\leq a \leq 1$ and which traces the fact that both $\Delta_t$ and $\Delta$ may tend to zero anisotropically. If $\tau=1$ and $\varepsilon \ll 1$ and $\Psi(t_j) \in C^2$, then Eq \ref{eq:evo1D} admits only a continuous limit in space time and it coincides with the following couple of PDEs, namely the Dirac equation in (1+1) spacetime dimensions:
\begin{equation}\label{eq:Dirac}
\partial_t \Psi= - \sigma_z \partial_x \Psi + \widehat C^{(1)}\Psi 
\end{equation}
if and only if $0< b <=1$ and 
\begin{equation}\label{eq:zeroAng}
\begin{split}
\theta^{(0)}   = 2q\pi \\
\alpha^{(0)}   = (q' + q")\pi \\
\xi^{(0)}   = (q' - q")\pi  
\end{split}
\end{equation}
and  $\forall ~\zeta^{(0)} ,~ \zeta^{(1)}, ~\theta^{(1)}  $ and where $q, q'$ and $q"$ $\in \mathbb{Z}$. 
\end{Th}
\textbf{Proof.} 
In order to prove the above Theorem, it is more suitable moving to Fourier space where the shift operator is diagonal. Using Lemma \ref{lem:FourierS} and Eq. \ref{eq:opFourier} we can then write the Eq. \ref{eq:evo1D} for $\tau=1$ as follows:
\begin{equation}\label{eq:Fourier1step}
\tilde \Psi(t+\Delta_t) =\widehat W_k \tilde  \Psi(t).
\end{equation}
where 
\begin{equation}
\widehat W_k =e^{i \alpha}\begin{pmatrix}e^{i (\xi-k\Delta)}\cos\theta & e^{i (\zeta-k\Delta)}\sin\theta \\ - e^{-i (\zeta-k\Delta)}\sin\theta & e^{-i (\xi-k\Delta)}\cos\theta \end{pmatrix}
\end{equation}
%Notice that we can gauge out the parameter $\xi$, with a simple change of variable $k  \rightarrow k  + \xi/\Delta$ which leaves the dynamics invariant. Then, for the following of the proof, we will set $\xi = 0$ without lack of generality. 
Notice that the vector $ \tilde \Psi(t) \in C^2$ because its Fourier inverse is. Now we can Taylor expand the above equation around $\varepsilon =0$ and up to the first order in $o(\varepsilon)$:
\begin{equation}
\tilde \Psi(t) (\mathds{1}- e^{- i \sigma_z k \varepsilon^{1-a}}\widehat C^{(0)}) = \varepsilon (\partial_t  + e^{- i \sigma_z k \varepsilon^{1-a}} \varepsilon^{b-1}\widehat C^{(1)}) \tilde \Psi(t) + o(\varepsilon^2)
\end{equation}
We can now distinguish two cases: (i) if $a$ is equal to 1 then $(\mathds{1}- e^{- i \sigma_z k}\widehat C^{(0)}) = 0$, implying that $\widehat C^{(0)} = e^{-i \sigma_z k}$ which is false because the coin does not depend on $k$. Thus, there not exist any continuous limit for $a=1$. (ii) If $0<a<1$, then at the zero$^{th}$ order, $\widehat C^{(0)}$ has to be the identity, which leads to Eqs. \ref{eq:zeroAng} after a simple and straightforward calculation. Thus, the leading order in $o(\varepsilon)$ reads: 
\begin{equation}
(\partial_t  -  \varepsilon^{-a}\sigma_z i k  + \varepsilon^{b-1}\widehat C^{(1)}) \tilde \Psi(t) + o(\varepsilon) = 0
\end{equation}
where $\widehat C^{(1)} $ is solely a function of $(\theta^{(1)}, \alpha^{(1)}, \xi^{(1)})$ and $\zeta$. Now, in order to avoid divergence $a = 0$ and $b$ have to be greater than or equal to the unity. Finally taking the limit $\varepsilon \rightarrow 0$ and inverse Fourier transform, we converge to:
\begin{equation}\label{eq: dirac1step}
\begin{split}
\partial_t \psi^0 &= e^{i\zeta} \theta^{(1)} \psi^1 - \partial_x \psi^0 + i (\alpha^{(1)} +\xi^{(1)})\psi^0,\\
\partial_t \psi^1 &= -e^{-i\zeta}\theta^{(1)} \psi^0 + \partial_x \psi^1 + i (\alpha^{(1)} -\xi^{(1)})\psi^1
\end{split}
\end{equation}
which coincides with Eq. \ref{eq:Dirac}, where $\widehat C^{(1)} = i \mathds{1} \alpha^{(1)} + i \sigma_z \xi^{(1)} - i(\sigma_y \cos(\zeta)- \sigma_x \sin(\zeta)) \theta^{(1)}$. 
and the $x-$, $y-$ and $z-$Pauli matrix read:
\begin{equation}\label{eq:Pauli}
\begin{split}
\sigma_x = \begin{pmatrix} 0&1\\1&0\end{pmatrix}\hspace{1cm}\sigma_y = \begin{pmatrix} 0&-i\\i&0\end{pmatrix}\hspace{1cm}\sigma_z = \begin{pmatrix} 1&0\\0&-1\end{pmatrix}.
\end{split}
\end{equation}

We have seen with the above Theorem that $a \neq 1$ precludes a continuous limit only in time (with discrete space) which means that there is no room for plasticity for $\tau=1$. Actually, this result is true for every odd $\tau$ and the reader will find detailed evidence in the recent \cite{manighalam2019continuum}. In the following, we will see that this problem can be encompassed considering $\tau=2$ (or more in general an even number of time steps). 

Starting from Eq. \ref{eq:1stepCont}, we can derive the stroboscopic equation of period $\tau =2 $

\begin{equation}\label{eq:2stepCont}
\Psi(t_j+2 \Delta_t) =  \widehat W^2 \Psi(t_j).
\end{equation}

As in the previous case, we will work in the Fourier basis and we will use the same truncated Taylor expansion for the quantum coin's angles given in \ref{eq:Cjet}.  \\

First we prove the following Lemma: 

\begin{Lem}[Zero$^{th}$ order]\label{lem:zero2}
The zero$^{th}$ order of the Taylor series of the stroboscopic equation \ref{eq:2stepCont}, for any value $0\leq a\leq 1$, cancels if and only if one of the following equations is satisfied:
\begin{enumerate}
\item $\forall a$, $\alpha^{(0)}   = \frac{\pi}{2} + q \pi $ and $ \theta^{(0)}   = \frac{\pi}{2} + q' \pi ~(C1)$
\item for $0\leq a< 1$, $\alpha^{(0)}   = \frac{\pi}{2} + q \pi $ and $ \xi^{(0)}   = \frac{\pi}{2} + q' \pi$ ~(C2)
\item for $0\leq a < 1$, $\alpha^{(0)}   = \frac{\pi}{2} q \pi $, ~ $ \xi^{(0)}   = \frac{\pi}{2} q' \pi$ and  $ \theta^{(0)}   = q" \pi$ ~(C3)
\end{enumerate}
where $q$, $q'$ and $q'' \in \mathbb{Z}$.
\end{Lem}
\textbf{Proof.}
Again we work in the Fourier basis. Then, let us Taylor expand \ref{eq:2stepCont} around $\varepsilon =0$ and at order zero we get:
\begin{equation}
\tilde{\Psi}(t_j) + o(\varepsilon) =  e^{-i \varepsilon^{1-a} \sigma_z k}\widehat C^{(0)}e^{-i \varepsilon^{1-a} \sigma_z k}\widehat C^{(0)} \tilde{\Psi}(t_j)  
\end{equation} 
The above equation is satisfied if 
\begin{equation}\label{eq:condzero}
e^{-i \varepsilon^{1-a} \sigma_z k}\widehat C^{(0)}e^{-i \varepsilon^{1-a} \sigma_z k}\widehat C^{(0)}  = \mathds{1},
\end{equation}
which reduces to:
\begin{small}
\begin{equation}
\cos^2(\varepsilon^{1-a} k)(\widehat C^{(0)})^2 - i \sin(\varepsilon^{1-a} k)\cos(\varepsilon^{1-a} k) \{\sigma_z,C^{(0)} \} - (\sigma_z C^{(0)})^2\sin^2(\varepsilon^{1-a} k)= \mathds{1}.
\end{equation}
\end{small}
Now we can distinguish two cases: (i) $a=1$ and (ii) $0\leq a <1$. Case (i) implies:
\begin{equation}
\cos^2(k)(\widehat C^{(0)})^2 - i \sin(k)\cos( k) \{\sigma_z,C^{(0)} \} - (\sigma_z C^{(0)})^2\sin^2(k)= \mathds{1}.
\end{equation}
From the above equation it follows that: 
\begin{equation}
\begin{split}
(\widehat C^{(0)})^2= \mathds{1}\\
\{\sigma_z,C^{(0)} \} = 0\\
(\sigma_z C^{(0)})^2 = -\mathds{1}.
\end{split}
\end{equation}
The first equation requires that $\widehat C^{(0)}$ be Hermitian, because $(\widehat C^{(0)})^2= \mathds{1} = \widehat C^{(0)} (\widehat C^{(0)})^{-1} $ implies $ (C^{(0)})^{-1} = C^{(0)}$. Notice that the last two conditions are equivalent. Both imply that $\widehat C^{(0)}$ be purely off diagonal, because the anti-commutation relations. Then, given $\widehat C^{(0)}$ as in \ref{def:coin}, the only solution is that $\widehat C^{(0)} = -\sigma_y$, where the $y$-Pauli matrix
\begin{equation}
\sigma_y = \begin{pmatrix}0 & -i \\ i & 0 \end{pmatrix}
\end{equation}
for any value of $\zeta$ and $\xi$, which directly implies $\alpha^{(0)}   = \frac{\pi}{2} + q \pi $ and $ \theta^{(0)}   = \frac{\pi}{2} + q' \pi$. 
Let us now prove the theorem for the conditions (ii). If $0\leq a <1$ is true then \ref{eq:condzero} reduces to:
\begin{equation}
\widehat C^{(0)}\widehat C^{(0)}  = \mathds{1},
\end{equation}
which is satisfied in the following three cases:
\begin{enumerate}
\item $\alpha^{(0)}   =\frac{\pi}{2} + q \pi$ and $\xi^{(0)}   =\frac{\pi}{2} + q' \pi$
\item $\alpha^{(0)}   =\frac{\pi}{2}q$, $\xi^{(0)}   =\frac{\pi}{2} q'$ and $\theta^{(0)}   = q" \pi$
\item $\alpha^{(0)}   =\frac{\pi}{2}+q\pi$, and $\theta^{(0)}   =\frac{\pi}{2} + q' \pi$
\end{enumerate}
which proves the Lemma.\\

%Notice that in Lemma \ref{lem:zero2}, there is no constraints on $\omega$. 

\begin{Rk}
Notice that from the above Lemma \ref{lem:zero2} it follows that for $a=1$ the continuous time limit in discrete space, for $\tau =2$,exists only if 
\begin{equation}
\begin{split}
\alpha^{(0)}   = \frac{\pi}{2} +q \pi \\
\theta^{(0)}   = \frac{\pi}{2} + q' \pi
\end{split}
\end{equation}
where $q$ and $q' \in \mathbb{Z}$.
\end{Rk}

Using \ref{lem:zero2} we are now able to prove the following Theorem: 

\begin{Th}[Continuous Limits for $\tau=2$]\label{th:CL1D_2steps}
Let the necessary conditions in Lemma \ref{lem:zero2} be verified, consequently the continuous limit of the QW in one spatial dimension, for a stroboscopic period $\tau = 2$, exists and reads:
%alpha=0
\begin{equation}\label{eq:theo2steps}
    \begin{split}
        \pmb{a=1:}\\
        \partial_t \Psi &= i \theta^{(1)} \sigma_y e^{-i \sigma_z (\zeta^{(0)}   + \xi^{(0)}  )} \left(\widehat S^2 + e^{-2i \sigma_z \xi^{(0)}  }\right)\Psi\\
      %  \partial_t \psi^0 &= e^{i \omega}\theta [e^{i \xi}\psi^1(x-2) + e^{-i \xi}\psi^1(x)] + 2 i \alpha^{(1)}   \psi^0(x)\\
        %\partial_t \psi^1 &= -e^{-i \omega}\theta [e^{-i \xi}\psi^0(x+2) + e^{i \xi}\psi^0(x)] + 2 i \alpha^{(1)}   \psi^1(x)\\
  \vspace{1cm}
        \pmb{0\leq a<1:}\\ 
         \partial_t \Psi &= P \partial_x \Psi + Q \Psi,
       % \partial_t \Psi^0 &= 2 e^{i \omega}\theta \cos(\xi)\psi^1 [e^{i \xi}\psi^1(x-2) + e^{-i \xi}\psi^1(x)] + 2 i \alpha^{(1)}   \psi^0(x)\\
        %\partial_t \psi^1 &= -e^{-i \omega}\theta [e^{-i \xi}\psi^0(x+2) + e^{i \xi}\psi^0(x)] + 2 i \alpha^{(1)}   \psi^1(x)
    \end{split}
\end{equation}
where $P$ and $Q$ are traceless and Hermitian matrices depending on $(\alpha^{(1)}, \xi^{(1)})$ and $\zeta$
\end{Th}
\textbf{Proof.}
Let us start with the case for which $a = 1$ and let us expand around $\varepsilon = 0$. In that case the Taylor truncated development of Eq. \ref{eq:2stepCont} reads:
\begin{equation}
\tilde{\Psi}(t_j)  + \varepsilon \partial_t \tilde{\Psi}(t_j) = e^{-i\sigma_z k}C^{(0)}e^{-i\sigma_z k}C^{(0)} + \varepsilon^{b} \{e^{-i\sigma_z k}C^{(1)},e^{-i\sigma_z k}C^{(0)}\}\tilde{\Psi}(t_j)  + o(\varepsilon^{2b}).
\end{equation}

From Lemma \ref{lem:zero2} the order$^{th}$ zero cancels and we get:
\begin{equation}
\partial_t \tilde{\Psi}(t_j)  =  \varepsilon^{b-1} \{e^{-i \sigma_z k}C^{(1)},e^{-i\sigma_z k}C^{(0)}\}\tilde{\Psi}(t_j)  + o(\varepsilon^{2b-1}),
\end{equation}
which constraint $b$ to be equal to 1 to avoid divergences. 
Taking the formal limit for $\varepsilon \to 0$, we recover the following equation in continuous time and discrete space:
\begin{equation}
\partial_t \tilde{\Psi}(t_j) =  \{e^{-i\sigma_z k}C^{(1)},e^{-i\sigma_z k}C^{(0)}\}\tilde{\Psi}(t_j) 
\end{equation}
using $\ref{lem:zero2}$ and Fourier transform we get
\begin{equation}
\partial_t \tilde{\Psi}(t_j) = i \theta \sigma_y e^{-i \sigma_z (\zeta^{(0)}   + \xi^{(0)}  )} \left(S^2 + e^{-2i \sigma_z \xi^{(0)}  }\right)\tilde{\Psi}(t_j) .
\end{equation}
proving the above Theorem for  $a = 1$.

Now let us explore the case for which  $a \neq 1$. The expansion of the Eq. \ref{eq:2stepCont} around $\varepsilon = 0$ leads to:
\begin{equation}
\tilde{\Psi}(t_j)  + \varepsilon \partial_t \tilde{\Psi}(t_j)  =\left( (\mathds{1} - i \sigma_z k \varepsilon^{1-a})(C^{(0)}+\varepsilon^b C^{(1)}) \right)^2 \tilde{\Psi}(t_j)  + o(\varepsilon^2).
\end{equation}
Now, after few simplifications and using the fact that the zero$^{th}$ order constraint equations \ref{eq:condzero} are satisfied, we get: 
\begin{equation}\label{eq:devEq1D}
\begin{split}
\partial_t \tilde{\Psi}(t_j)  &= (- i k \varepsilon^{-a} \{\sigma_z C^{(0)}, C^{(0)}\} + \varepsilon^{b-1}\{C^{(0)},C^{(1)}\} \\
&- i \varepsilon^{-a+b} k [\{\sigma_z C^{(0)}, C^{(1)}\}+\{\sigma_z C^{(1)}, C^{(0)}\}])\tilde{\Psi}(t_j)  + o(\varepsilon^{2(1-a)-1}) +o(\varepsilon^{2b-1}).
\end{split}
\end{equation}
From the above equation we deduce that if  $\{\sigma_z C^{(0)}, C^{(0)}\} \neq 0$, $a=0$ to avoid divergences. Consequently, we will study the coefficients of the above development for each case of Lemma \ref{lem:zero2}. Let us start with C1: in this case $\{\sigma_z C^{(0)}, C^{(0)}\} = 0$ then in order to recover the spatial derivative we need $a=b$. Because $0\leq a<1$, $\{C^{(0)},C^{(1)}\}$ has to be 0, otherwise $b-1$ would be negative and we would have a divergence in the limit for $\varepsilon \rightarrow 0$. This implies the supplementary condition:
\begin{enumerate}
\item $ \alpha^{(1)} = 0$ and $\theta^{(1)}= 0$,
\item $ \alpha^{(1)}= 0$ and $\xi^{(0)}   = \pm\frac{\pi}{2}$.
\end{enumerate}
The second one of the above conditions is the only one for which $[\{\sigma_z C^{(0)}, C^{(1)}\}+\{\sigma_z C^{(1)}, C^{(0)}\}] \neq 0$.  
Plugging it in \ref{eq:devEq1D} together with \ref{lem:zero2}.C1, together with the formal limit for $\varepsilon \to 0$, leads us to:
\begin{equation}\label{eq:strong}
\begin{split}
\partial_t \psi^0 &= 2 e^{i \zeta} \theta^{(1)} (\xi^{(1)}  \psi^{1} + i \partial_x \psi^{1}),\\
\partial_t \psi^1 &= -2 e^{-i \zeta}\theta^{(1)}  (\xi^{(1)}  \psi^{0} + i \partial_x \psi^{0}).
\end{split}
\end{equation}
Notice that the above equations are the continuous space limit of Eqs. \ref{eq:theo2steps}.a. \\

Now let us consider the condition C2 in the Lemma \ref{lem:zero2}, for which $\alpha^{(0)}   = \pi/2$ and $\xi^{(0)}   = \pi/2$. For these values $\{\sigma_z C^{(0)},C^{(0)}\} \neq 0$, thus we can explore separately the case for $a =0$. The Eq. \ref{eq:devEq1D} becomes:
\begin{equation}\label{eq:devEq1D}
\begin{split}
\partial_t \tilde{\Psi}(t_j) &= (- i k \{\sigma_z C^{(0)}, C^{(0)}\} \tilde{\Psi}(t_j) + \varepsilon^{b-1}\{C^{(0)},C^{(1)}\} \tilde{\Psi}(t_j)  + o(\varepsilon^{1}) +o(\varepsilon^{b})
\end{split}
\end{equation}
and for $b=1$, it follows, after Fourier transform:

\begin{equation}
\begin{split}\label{eq:weak1}
\partial_t \psi^0 &= e^{i \zeta} \sin(2 \theta) (i \partial_x \psi^1 +  \xi^{(1)}  \psi^1) + 2 i \cos^2(\theta) (i \partial_x u + \xi^{(1)}  \psi^0) + 2i \alpha^{(1)} \psi^0\\
\partial_t \psi^1 &= -e^{-i \zeta} \sin(2 \theta) (i \partial_x \psi^0 + \xi^{(1)}  \psi^0) - 2 i \cos^2(\theta) (i \partial_x 1 + \xi^{(1)}  \psi^1) + 2i\alpha^{(1)}  \psi^1 .
\end{split}
\end{equation}

The case of $a < 1$ requires that $\{\sigma_z C^{(0)},C^{(0)}\} = 0$, which leads to $\theta^{(0)}   = \pm \pi/2$, coming back to \ref{lem:zero2}.C1. \\

The last case, \ref{lem:zero2}.C3 has to be split in two sub-cases: (a) $q=q'=1$ and $q"=0$ that is $\alpha^{(0)}   = \xi^{(0)}   = \frac{\pi}{2} \pi$ and  $ \theta^{(0)}   = 0$ and (b) $q=q'=0$ and $q"=0$ that is $\alpha^{(0)}   =\xi^{(0)}   = \theta^{(0)}   = 0$. For $a=0$ we obtain in inverse Fourier basis:
\begin{equation}
\begin{split}\label{eq:weak2}
\partial_t \psi^0 &= -2\partial_x \psi^0 +2i (\alpha^{(1)}  + \xi^{(1)} )\\
\partial_t \psi^1 &= 2\partial_x \psi^1  +  2i (\alpha^{(1)}  - \xi^{(1)} )
\end{split}
\end{equation}
Notice that there is no limits for $0<a <1$ because for $\theta^{(0)}   = 0$, $\{\sigma_z C^{(0)},C^{(0)}\}$ is always non vanishing. \\
Case (b) leads to 
\begin{equation}
\begin{split}\label{eq:weak2}
\partial_t \psi^0 &= -2\partial_x \psi^0 +2i ( \alpha^{(1)}  + \xi^{(1)} ) + 2 e^{i\zeta}\theta^{(1)}   \psi^1\\
\partial_t \psi^1 &= 2\partial_x \psi^1  +  2i (\alpha^{(1)}  - \xi^{(1)} )- 2 e^{-i\zeta}\theta^{(1)}   \psi^0
\end{split}
\end{equation}
As in the previous case $\theta^{(0)}   = 0$ will make $\{\sigma_z C^{(0)},C^{(0)}\}$ always non vanishing. Finally all Eqs \ref{eq:strong}, \ref{eq:weak1} and \ref{eq:weak2} can be recasted as $\partial_t \Psi = P \partial_x \Psi + Q$ where $P$ and $Q$ are Pauli matrices or combinations of them and then Hermitian and traceless matrices. \\

In conclusion we can resuming the above Lemmas and Theorems, with the following Theorem distinguish three families of PQW: 

\begin{Th}[Plastic Quantum Walks in 1D]
If $\tau=2$ and $d=1$ and one of the following conditions is satisfied 
\begin{enumerate}
\item $\alpha^{(0)}   = \frac{\pi}{2} + q \pi $ and $ \theta^{(0)}   = \frac{\pi}{2} + q' \pi$, $ \xi^{(0)}   = \frac{\pi}{2} + q" \pi$, $\alpha^{(1)}   =0$ ~$[\mathcal{S_1}]$
\item $\alpha^{(0)}   = \frac{\pi}{2} + q \pi $, $ \xi^{(0)}   = \arccos(\cos(\xi') \varepsilon^{1-\alpha})$, and $ \theta^{(0)}  = \arccos(\cos(\theta') \varepsilon^{\alpha})$ ~$[\mathcal{P}'_{1a}]$
\item $\alpha^{(0)}   = \frac{\pi}{2} + q \pi $, $ \xi^{(0)}   = \frac{\pi}{2} q"$, and $ \theta^{(0)}  = \arccos(\cos(\pi q') \varepsilon^{\alpha})$ ~$[\mathcal{P}"_{1a}]$
\end{enumerate}
where $\xi'$ and $\theta'  \in [0,2\pi]$, then the QW is Plastic. In particular we will refer to the first PQW as that one with \textit{strong plasticity}. 
%This follows from the fact that this class of PQW is the only one for which the zero$^{th}$ conditions do not change with $a$. 
\end{Th}

\begin{Def}[Strong Plastic Quantum Walk]
A Strong PQW, $\mathcal{S}_{1}$, is a QW which admits \textbf{exactly} both a continuous limit in time and a continuous limit in spacetime.
\end{Def}
where by ``exactly" we mean that the constraint conditions at zero$^{th}$ order do not depend on $a$ and are exactly the same for the continuous time and continuous spacetime limit. Such a plasticity is more powerful because the QW does not need to be adapted to admit both limits, \ie the coin will be kept the same, and the convergence to each regime will be solely a property of the configuration space itself, the anisotropy of the grid. 

We will call the other classes of PQW, $\mathcal{P}$, weak PQW because the constraint conditions depends on $a$, which means that to converge in both formal limits (continuous spacetime and continuous time and discrete space) we need to engineer a different coin which will scale with the anisotropy parameter $a$ of the grid. In the following we will not consider them. 

\section{Convergence}

In numerical analysis, in order to evaluate the quality of a numerical scheme model, the most important criterion for quality is \textit{convergence}. From an intuitive point of view it requires that, after an arbitrary time $x_0$, and if $\varepsilon$, the discretisation step, has been chosen small enough, the discrete model approximates the solution to a given order of $\varepsilon$. Arrighi, Nesme, Forets \cite{ArrighiDirac}, proved rigorously that the probability of observing a discrepancy between the iterated quantum walk and the analytical solution of the Dirac equation in arbitrary dimension converge, \textit{i.e.}, goes to zero, quadratically as the discretisation step $\varepsilon$ goes to zero. The convergence was already argued in \cite{meyer1996quantum} and numerically in \cite{love2005dirac}. Di Molfetta and Debbasch \cite{Dimolfetta2014aa} proved numerically the convergence in a case where the operator was neither invariant for translations nor homogeneous. To ensure convergence, we will assume throughout the entire manuscript that $\varepsilon$ could be always chosen arbitrarily small.

\section{Discussion and open problems}

In this chapter we introduced the concept of plasticity. This characteristic reveals the fundamental property of different QWs to adapt to the anisotropy of the space-time grid. This elasticity, in its strong version, makes the QW itself able to formally follow both limits, in continuous time and discrete space time. It offers a class of QWs that have the ability to unify in a single mathematical representation, the main simulation schemes in both continuous time (Hamiltonian evolutions) and in discrete time (QWs or more generally QCA). Although such a result already lends itself to a wide range of applications that we will discuss in the conclusion of the thesis, there is still much to do, certainly in at least three directions: (i) the generalization of the results obtained in 1D to higher dimensions; (ii) an extension of the concept of plasticity to arbitrary simplicial complex. In fact Plastic QWs is likely to prove more powerful in a space where not only there is anisotropy, but the very concept of grid is missing, suggesting a coupling between plasticity and the intrinsic topology of the simplicial complexes. (iii) Finally, there is the question of symmetries: it is well known that continuous time computation models, although widely used, often suffer from the violation of some important symmetries in nature such as Lorentz- covariance; the bounded speed of light can only be approximately recovered e.g. via Lieb-Robinson bounds. This also creates more subtle problems such as \textit{fermion doubling}, where spurious particles are created due to the periodic nature of the momentum space on a lattice. This latter problem is not present in QW. It remains therefore to be investigated whether PQW can finally solve such problems in continuous time, inheriting the good properties of QW in discrete time. In this thesis we will not develop any of these points that we will leave for future research. Instead, in Chapter 5 we will lay the foundations to investigate the second point, i.e. we investigate QWs over simplicial complexes, on arbitrary triangulations. 

\medskip

%We have discussed the various problems related to the family of definitions $QX$, and so we ought to do the same with definitions $QZ$, except there is none of the previously mentioned problems with this family of definitions: they are operational, unitary, causal. The issue however is one of generality. Apart from in \cite{SchumacherWerner}, these definitions appear to be quite ad hoc at first:  Why should we believe that QCA in general are of the particular form introduced by \cite{Watrous}, \cite{VanDam}, \cite{BrennenWilliams}, \cite{NagajWocjan}, \cite{Raussendorf} or \cite{PerezCheung}? Is it the case at all that all QCA admit a block representation? As we have explained in Subsection \ref{subsec:y} this is not even always the case in the classical case. We postpone the answers to these questions to Chapter \ref{chap:structures}.

 %14 pages -- proofread
%14 pages

\chapter{Plasticity in 2D, transport equations and beyond}\label{chap:Continuous Limit2D}

%\toquotebig{CITE}

\toabstract{
We will prove the existence of Plastic Quantum Walks on the square grid. Surprisingly as a byproduct of the very general method we introduce, we will show how some of them admit as limit several families of PDE encompassing both transport equations and dispersive linear partial difference equations, namely the Schr\"odinger equation. These results pave the way to simulate quantum transport in dispersive media.}

\clearpage

\section{Higher spatial dimensions: the plane}\label{eq:2D}

The higher dimensional case is more complicate and analytical calculations are less straightforward. This is why we will try to operate several simplifications without lack of generality. First we can observe that proving the strong plasticity of a QW do not require any constraints over the first order of the angles except for $\alpha^{(1)}$ which needs to be $0$. We can also remark that the parameter $\theta^{(1)}$ plays a fundamental role in the continuous time limit because it is the pre-factor of the displacement operators. Although this is certainly true in 1D, we conjecture that this has to be true also in 2D and we will prove a posteriori that it will be the case. This conjecture will lead us to consider only the zero$^{th}$ order of the jet proposed in \ref{eq:scalings} except for $\theta$. \\
We start from the same AQW, defined in \ref{def:alternateQW} in two spatial dimensions, where the coin driving the walker along each spatial dimensions is now in general a different one. Moreover, in order to further simplify the calculations, we will write each coin as follows:

\begin{equation}\label{eq:rotations}
    \begin{split}
    C_j&=e^{i\alpha_j}R_z(\omega)R_y(\nu)R_z(\phi)=e^{i\alpha_j}e^{ -i\omega\sigma_z/2}e^{ -i\nu\sigma_y/2} e^{ -i\phi\sigma_z/2}
        \end{split}
\end{equation}
where
\begin{equation}
\begin{split}
\phi_j=(\xi_j+\zeta_j)/2 \\
\omega_j = (\xi_j-\zeta_j)/2 \\
\nu_j = -2\theta_j
\end{split}
\end{equation}
and $j = x,y$.

Again we investigate for each period $\tau$ and in particular for $\tau= 1$ and $\tau = 2$ whether we recover plasticity. As in the one dimensional case we will see that there is no room for plasticity for $\tau = 1$. In order to prove it, we will start to investigate for which $\tau$ there exists the continuous time limit. 

Using Eq.\ref{eq:rotations} and the scalings \ref{eq:scalings}, and expanding around $\varepsilon = 0$, the rotation matrices $R_m(w)$ read:
\begin{equation}
    R_m(w)\simeq R_m(w_0)(1-\frac{i w^{(1)} \varepsilon}{2}\sigma_m+o(\varepsilon^2)),
\end{equation}
 where $w=\omega$,$\nu$,$\phi$ and $m=x,y$. 
Because we are interested to find the sufficient conditions to recover the limit at first order $o( \varepsilon)$, as in the previous section, we will study the zero$^{th}$ order, \ie for $\varepsilon=1$. The zero$^{th}$ order of the alternated unitary operator:

\begin{equation}\label{eq:w}
    \begin{split}
    \widehat{W}\simeq e^{i\alpha^{(0)}  }A,
    \end{split}
\end{equation}
where $A=A_xA_y$, $\alpha^{(0)}   = \alpha_x^{(0)}  +\alpha_y^{(0)}  $ and 
\begin{equation}
\begin{split}
 A_j&=R_z(\omega_j^{'(0)}  )R_y(\nu_{j}^{(0)})R_z(\phi_{j}^{(0)})\\
\omega_j^{'(0)}  &=\omega_j^{(0)}  +2k_j.
\end{split}
\end{equation}
Finally, in order to compute the zero orders of Eq. \eqref{eq:w} for $\tau = 0$ and $\tau = 1$, let us compute the $\tau^{th}-$power of $W$:
\begin{equation}\label{eq:reduced1sc^n}
    \begin{split}
    \widehat{W}^\tau&\simeq e^{i\alpha^{(0)}   \tau}A^\tau =(e^{i\alpha^{(0)}  }A)^\tau.
    \end{split}
\end{equation}

Now we have the following three Lemmas:
 
\begin{Lem}\label{lma:n=1}
An AQW in two spatial dimensions does not admit a continuous time limit for $\tau=1$.
\end{Lem}
\textbf{Proof.}
In order to satisfy the continuous limit $\widehat{W}^\tau$ must equal $\mathds{1}$, and thus $\widehat{W}$ must equal $\mathds{1}$ as well. Therefore, from Eq.~\eqref{eq:reduced1sc^n}, $(e^{i\alpha^{(0)}  }A)^\tau$ must equal identity if $\widehat{W}=\mathds{1}$. The only unitary operator $e^{i\alpha^{(0)}  }A$ that could possibly satisfy $(e^{i\alpha^{(0)}  }A)^\tau=\mathds{1}$ for $\tau=1$ is the identity operator itself. But $e^{i\alpha^{(0)}  }A$ cannot even equal identity, as $A$ has $k_x$ and $k_y$ dependence from containing $\widehat{S}_{u_x}$ and $\widehat{S}_{u_y}$, and the angles are not permitted to depend on $k_x$ and $k_y$, so there is no possible way to cancel out the $k_x$ and $k_y$ dependence. Thus, there is no continuous time limit for a two-dimensional AQW for $\tau=1$.\\

The above results confirm the one-dimensional case. Now let us move to the case $\tau=2$.
 
\begin{Lem}\label{lma:constraints}
For the continuous time limit to exist, $|(\nu_x^{(0)}  -\nu_{y}^{(0)})\text{ mod }2\pi|=\pi$, and $\alpha^{(0)}  =2\pi l-\frac{p\pi}{2}$ for odd integer $p$ and for any positive integer number $l$.
\end{Lem}
\textbf{Proof.}
Following up on the constraint that $(e^{i\alpha^{(0)}  }A)^2=\mathds{1}$ from Eq. \eqref{eq:reduced1sc^n}, let $U$ be the diagonalization matrix of $A$, and let $D$ be the matrix of eigenvalues of $A$. Then we have the following:
\begin{align}
    &(e^{i\alpha^{(0)}  }A)^2=e^{2i\alpha^{(0)}  }U^{-1}D^2 U=\mathds{1} \rightarrow e^{2i\alpha^{(0)}  }D^2=
    UU^{-1}=\mathds{1}\rightarrow e^{2i\alpha^{(0)}  }D^2=\mathds{1}
\end{align}
 so if we set the eigenvalues of $e^{i\alpha^{(0)}  }A$ equal to a $2^{nd}$ root of unity $e^{\pi i l}$ where $l=0,1,2,..$ (which is equivalent to the constraint $(e^{i\alpha}A)^2=\mathds{1}$), we recover the following constraint equation for $\nu_x^{(0)}  $ and $\nu_{y}^{(0)}$:
\begin{equation}\label{eq:constraint_function_simple}
    f(k_x,k_y)=a\cos(g(k_x,k_y))-b\cos(h(k_x,k_y))-c=0
\end{equation}
where 
\begin{equation}
\begin{split}
a&=\cos(\frac{\nu_x^{(0)}  }{2})\cos(\frac{\nu_{y}^{(0)}}{2}),\\
b&=\sin(\frac{\nu_x^{(0)}  }{2})\sin(\frac{\nu_{y}^{(0)}}{2}),\\
c&=\cos(\pi l-\alpha^{(0)}),\\
g(k_x,k_y)&=\frac{\phi_x^{(0)}  +\phi_y^{(0)}  +\omega_{x}^{'(0)}(k_x)+\omega_{y}^{'(0)}(k_y)}{2},\\
h(k_x,k_y)&=\frac{\phi_y^{(0)}  -\phi_x^{(0)}  +\omega_{x}^{'(0)}(k_x)-\omega_{y}^{'(0)}(k_y)}{2}.
\end{split}
\end{equation}

Notice that the above constraint equation has to hold for all $k_x$ and $k_y$ and additionally, all derivatives of $f(k_x,k_y)$ with respect to $k_x$ and $k_y$ must equal zero as well. Using $\frac{\partial g(k_x,k_y)}{\partial k_x}=-1$, $\frac{\partial h(k_x,k_y)}{\partial k_x}=-1$, $\frac{\partial g(k_x,k_y)}{\partial k_y}=-1$, and $\frac{\partial h(k_x,k_y)}{\partial k_y}=1$ we obtain the derivative of $f(k_x,k_y)$ with respect to $k_x$ and $k_y$:
\begin{equation}
    \begin{split}
        &\frac{\partial f(k_x,k_y)}{\partial k_x}=a\sin(g(k_x,k_y))-b\sin(h(k_x,k_y))=0\\
        &\frac{\partial f(k_x,k_y)}{\partial k_y}=a\sin(g(k_x,k_y))+b\sin(h(k_x,k_y))=0.
    \end{split}
\end{equation}
For both of these equations to be true, we must have the following:
\begin{equation}
    \begin{split}
        &a\sin(g(k_x,k_y))=0\\
        &b\sin(h(k_x,k_y))=0.
    \end{split}
\end{equation}
Due to $\phi_{i}^{(0)}$ and $\omega_{i}^{(0)}$ being parameters which cannot depend on $k_{i}$, it follows that $\sin(g(k_x,k_y))$ cannot equal zero for all values of $k_x$ and $k_y$, so the following must be true:
\begin{equation}\label{eq:constraints}
    \begin{split}
        &a=0\rightarrow\cos(\frac{\nu_x^{(0)}  }{2})\cos(\frac{\nu_{y}^{(0)}}{2})=0\rightarrow \nu_{i}^{(0)}=2q\pi+\pi \text{ for any integer $q$, and $i=x$ or $y$}\\
        &b=0\rightarrow\sin(\frac{\nu_x^{(0)}  }{2})\sin(\frac{\nu_{y}^{(0)}}{2})=0\rightarrow \nu_{j}^{(0)}=2\pi r \text{ for any integer $r$, and $j\neq i$}
    \end{split}
\end{equation}
In other words, $|(\nu_x^{(0)}  -\nu_{y}^{(0)})\text{ mod }2\pi|=\pi$. This corresponds to either $C_x$ purely diagonal and $C_y$ purely off-diagonal, or vice-versa. Further, because $a,~b=0$, it must be true from Eq.~\eqref{eq:constraint_function_simple} that $c=0\rightarrow \cos(\pi l-\alpha^{(0)}  )=0\rightarrow \alpha^{(0)}  =\pi l-\frac{p\pi}{2}$ for odd integer $p$ and any positive integer number $l$. Note that this condition encompass that one we got in 1D over $\alpha^{(0)}$. 
  
\begin{Lem}\label{lma:CTLimitH}
Let $\nu_x^{(0)}  =2\pi m+\lambda \pi$ and $\nu_{y}^{(0)}=2\pi t+(1-\lambda)\pi$, where $\lambda \in \mathbb{B}$ parametrizes the constraints in Eq.~\eqref{eq:constraints}. The continuous time limit of an AQW in two spatial dimensions is:
%alpha=0
\begin{equation}
    \begin{split}
        %\pmb{\lambda=0:}\\
        \partial_t \Psi =\frac{-i }{4}\big[&\nu_{x}^{(1)}\big(S_{u_x}^2R_z(\omega_x^{(0)}  )+S_{u_y}^{(-1)^{\lambda}2}R_z((-1)^{\lambda}2\omega_y^{(0)} +(-1)^{\lambda}2\phi_x^{(0)}  -2\phi_y^{(0)}  )\big)\\
        +&\nu_{y}^{(1)}\big(R_z(-2\phi_y^{(0)}  )+S_{u_x}^2S_{u_y}^{(-1)^{\lambda}2}R_z(2\omega_x^{(0)}+ (-1)^{\lambda}2\phi_x^{(0)}  +(-1)^{\lambda}2\omega_y^{(0)}  )\big)\big]\sigma_y \Psi
    \end{split}
\end{equation}
%
%%alpha=1
%\begin{equation}
%    \begin{split}
%        \pmb{\lambda=1:}\\
%     \hspace{0.9cm}    \partial_t \Psi =\frac{-i}{4}\big[&\nu_{x}^{(1)}\big(S_{u_x}^2R_z(\omega_x^{(0)}  )+S_{u_y}^{-2}R_z(-2\omega_y^{(0)}  -2\phi_x^{(0)}  -2\phi_y^{(0)}  )\big)\\
%        +&\nu_{y}^{(1)}\big(R_z(-2\phi_y^{(0)}  )+S_{u_x}^2S_{u_y}^{-2}R_z(2\omega_x^{(0)}  -2\phi_x^{(0)}  -2\omega_y^{(0)}  )\big)\big]\sigma_y  \Psi
%    \end{split}
%\end{equation}
%
\end{Lem}
\textbf{Proof.}
We begin by using that $A^2=-1$, $A^{-1}=-A$, and $(e^{i\alpha^{(0)}}A)^2=\mathds{1}$ to reduce Eq.~\eqref{eq:reduced1sc^n}: 
\begin{equation}\label{eq:reduced1sc^n_2}
    \begin{split}
    \widehat{W}^2&=e^{2 i\alpha^{(0)}}(A-\frac{i \varepsilon}{2} B)^2\\
    &=(e^{i\alpha^{(0)}}A)^2(\mathds{1}-\frac{i \varepsilon}{2}A^{-1}\sum_{j=0}^{1}A^{-j}BA^j+o(\varepsilon^2))\\
    &=\mathds{1}+\frac{i \varepsilon}{2}A\sum_{j=0}^{1}(-1)^jA^{j}BA^j+o(\varepsilon^2).
    \end{split}
\end{equation}
Now we evaluate the sum:
\begin{equation}
    \begin{split}
        A\sum_{j=0}^{1}(-1)^j A^{j}BA^j&=A(-ABA+B)=\{A,B\}.
    \end{split}
\end{equation}

After a tedious but straightforward calculations, we have the following for $\{A,B\}$:
\begin{equation}
    \begin{split}
        \{A,B\}=&-\nu_{y}^{(1)}(R_z(-2\phi_y^{(0)}  )+R_z(2\omega_{x}^{'(0)}+2\phi_x^{(0)}  (-1)^\lambda+2\omega_{y}^{'(0)}(-1)^\lambda))\sigma_y\\
        &-\nu_{x}^{(1)}(R_z(2\omega_{x}^{'(0)})+R_z(2\omega_{y}^{'(0)}(-1)^\lambda-2\phi_y^{(0)}  +2\phi_x^{(0)}  (-1)^\lambda))\sigma_y
    \end{split}
\end{equation}
Now we have the following for Eq.~\eqref{eq:reduced1sc^n_2}:
\begin{equation}
    \begin{split}
        \widehat{W}^2&=\mathds{1}+\frac{i \varepsilon}{2}A\sum_{j=0}^{1}(-1)^jA^{j}BA^j+o(\varepsilon^2)\\
        &=\mathds{1}+\frac{i\varepsilon}{2}\{A,B\}+o(\varepsilon^2)\\
        &=\mathds{1}-\frac{i\varepsilon}{2}(\nu_{y}^{(1)}(R_z(-2\phi_y^{(0)}  )+R_z(2\omega_{x}^{'(0)}+2\phi_x^{(0)}  (-1)^\lambda+2\omega_{y}^{'(0)}(-1)^\lambda))\\
        &+\nu_{x}^{(1)}(R_z(2\omega_{x}^{'(0)})+R_z(2\omega_{y}^{'(0)}(-1)^\lambda-2\phi_y^{(0)}  +2\phi_x^{(0)}  (-1)^\lambda)))\sigma_y+o(\varepsilon^2)
    \end{split}
\end{equation}
And finally, taking formally the limit for $\varepsilon \to 0$ and converting to real space, we get the following
\begin{small}
\begin{equation}\label{eq:CTLimit_H_alpha}
    \begin{split}
        \partial_t\Psi=\frac{1}{4}\big[&\nu_{x}^{(1)}\big(S_{u_x}^2R_z(\omega_x^{(0)}  )+S_{u_y}^{2(-1)^\lambda}R_z(2\omega_y^{(0)}  (-1)^\lambda+2\phi_x^{(0)}  (-1)^\lambda-2\phi_y^{(0)}  )\big)\\
        +&\nu_{y}^{(1)}\big(R_z(-2\phi_y^{(0)}  )+S_{u_x}^2S_{u_y}^{2(-1)^\lambda}R_z(2\omega_x^{(0)}  +2\phi_x^{(0)}  (-1)^\lambda+2\omega_y^{(0)}  (-1)^\lambda)\big)\big]\sigma_y  \Psi  
    \end{split}
\end{equation}
\end{small}
%Notice that $\phi^{(1)}$ and $\omega^{(1)}$ don't appear in the above equation, only $\nu_{x}^{(1)}$ is relevant. To further simplifications for the next part, we will set both parameters to zero.

\begin{Rk}
Notice that, differently from the one dimensional case, in the two dimensional case when $\tau=2$ we recover cross derivatives. We will see that this is also the case in the continuous spacetime limit. Now imagine to reduce the above 2D AQW in 1D. The same QW would coincide with a one dimensional QW with inhomogeneous coin and with $\tau=4$, then the cross derivatives becomes second order derivatives in space. The PDE will be no longer hyperbolic because the presence of the parabolic terms. We may argue that this will happen for any power of $4$ and we believe that this will deserve further investigations. 
\end{Rk}

Now we want to find out all possibles families of AQWs, admitting a continuous time limit as shown in Lemma \ref{lma:CTLimitH}, which also admit a continuous space-time limit. The aim is looking for strong PQW in 2D. As in the one dimensional case we parametrize the space steps as follows:
\begin{equation}
    \Delta_x=\Delta_y=\Delta = \varepsilon^{(1-a)}
\end{equation}
where $a\in[0,1)$. 
Now we expand the operator $\widehat S_{u_m}\widehat C_m$ in powers of $\varepsilon$, where $m=x,y$:
\begin{equation}
    \begin{split}
        \widehat S_{u_m}\widehat C_m&=e^{\alpha^{(0)}_m}R_z(\omega^{'(0)}_{m})R_y(\theta_{m})R_z(\phi^{(0)}_{m})\\
        &=e^{\alpha^{(0)}_m}R_z(\omega^{'(0)}_{m})\sum_{n_m=-\infty}^{\infty}\frac{(-\frac{i\theta_{m}^{(1)}\sigma_y}{2})^{n_m}}{n_m!}\varepsilon^{n_mb}R_y(\theta_{m}^{(0)})R_z(\phi^{(0)}_{m})\\
        &=e^{\alpha^{(0)}_m}R_z(\omega^{(0)}_{m})\sum_{l_m=-\infty}^{\infty}\frac{(ik_m\sigma_z)^{l_m}}{l_m!}\varepsilon^{l_m(1-a)}\sum_{n_m=-\infty}^{\infty}\frac{(-\frac{i\theta_{m}^{(1)}\sigma_y}{2})^{n_m}}{n_m!}\varepsilon^{n_mb}R_y(\theta_{m}^{(0)})R_z(\phi_{m}^{(0)})\\
        &=e^{\alpha^{(0)}_m}\sum_{l_m,n_m}\left(\varepsilon^{l_m(1-a)+n_mb}\frac{(ik_m)^{l_m}(-\frac{i\theta_{m}^{(1)}}{2})^{n_m}}{l_m!n_m!}R_z(\omega_{m}^{(0)})\sigma_z^{l_m}\sigma_y^{n_m}R_y(\theta_{m}^{(0)})R_z(\phi_{m}^{(0)})\right)
    \end{split}
\end{equation}
Next we use the above equation to expand $\widehat S_{u_x}\widehat C_x\widehat S_{u_y}\widehat C_y$ in powers of $\varepsilon$:
\begin{equation}
    \begin{split}
        \widehat S_{u_x}\widehat C_x\widehat S_{u_y}\widehat C_y=&e^{2i\alpha^{(0)}}\sum_{\substack{l_x,n_x\\l_y,n_y}}\varepsilon^{(1-a)(l_x+l_y)+b(n_x+n_y)}\frac{(ik_x)^{l_x}(ik_y)^{l_y}(-\frac{i\theta_{x}^{(1)}}{2})^{n_x}(-\frac{i\theta_{y}^{(1)}}{2})^{n_y}}{l_x!l_y!n_x!n_y!}\\
        &\times R_z(\omega_{x}^{(0)})\sigma_z^{l_x}\sigma_y^{n_x}R_y(\nu_{x}^{(0)})R_z(\phi_{x}^{(0)})R_z(\omega_{y}^{(0)})\sigma_z^{l_y}\sigma_y^{n_y}R_y(\nu_{y}^{(0)})R_z(\phi_{y}^{(0)})\\
        &=e^{2i\alpha^{(0)}}\sum_{\substack{l_x,n_x\\l_y,n_y}}\varepsilon^{(1-a)(l_x+l_y)+b(n_x+n_y)}\Lambda_{l_xl_yn_xn_y}\widehat{\Gamma}_{l_xl_yn_xn_y},
    \end{split}
\end{equation}
where
\begin{equation}
    \Lambda_{l_xl_yn_xn_y}=\frac{(ik_x)^{l_x}(ik_y)^{l_y}(-\frac{i\nu_{x}^{(1)}}{2})^{n_x}(-\frac{i\nu_{y}^{(1)}}{2})^{n_y}}{l_x!l_y!n_x!n_y!}
\end{equation}
and
\begin{equation}\label{eq:unitaryCoeffs}
    \begin{split}
        \widehat{\Gamma}_{l_xl_yn_xn_y}&=R_z(\omega_{x}^{(0)})\sigma_z^{l_x}\sigma_y^{n_x}R_y(\nu_{x}^{(0)})R_z(\phi_{x}^{(0)})R_z(\omega_{y}^{(0)})\sigma_z^{l_y}\sigma_y^{n_y}R_y(\nu_{y}^{(0)})R_z(\phi_{y}^{(0)})
        %&=\sigma_z^{l_x+l_y}\sigma_y^{n_x+n_y}R_z[(-1)^{n_x+n_y}\omega_{x}^{(0)}]R_y[(-1)^{l_y}\nu_{x}^{(0)}]\\
        %&\times R_z[(-1)^{n_y}(\phi_{x}^{(0)}+\omega_{y}^{(0)})]R_y[\nu_{y}^{(0)}]R_z[\phi_{y}^{(0)}]
    \end{split}
\end{equation}
Now we have the following for $(\widehat S_{u_x}\widehat C_x\widehat S_{u_y}\widehat C_y)^2$:
\begin{equation}\label{eq:(sxcxsycy)^2}
    \begin{split}
        (\widehat S_{u_x}\widehat C_x\widehat S_{u_y}\widehat C_y)^2&=e^{2i\alpha^{(0)}}(\sum_{\substack{l_x,n_x\\l_y,n_y}}\varepsilon^{(1-a)(l_x+l_y)+b(n_x+n_y)}\Lambda_{l_xl_yn_xn_y}\widehat{\Gamma}_{l_xl_yn_xn_y})^2\\
        &=e^{2i\alpha^{(0)}}\sum_{\substack{l_{1x},l_{1y}\\l_{2x},l_{2y}}}\sum_{\substack{n_{1x},n_{1y}\\n_{2x},n_{2y}}}\varepsilon^{(1-a)(l_{1x}+l_{1y}+l_{2x}+l_{2y})+b(n_{1x}+n_{1y}+n_{2x}+n_{2y})}\\
        &\times \Lambda_{l_{1x}l_{1y}n_{1x}n_{1y}}\Lambda_{l_{2x}l_{2y}n_{2x}n_{2y}}\widehat{\Gamma}_{l_{1x}l_{1y}n_{1x}n_{1y}}\widehat{\Gamma}_{l_{2x}l_{2y}n_{2x}n_{2y}}\\
        &=e^{2i\alpha^{(0)}}\Lambda_{0000}^2\widehat{\Gamma}_{0000}^2\\
        &+e^{2i\alpha^{(0)}}\sum_{\substack{l_{1x},l_{1y}\\l_{2x},l_{2y}\\\neq (0,0,0,0)}}\sum_{\substack{n_{1x},n_{1y}\\n_{2x},n_{2y}\\\neq (0,0,0,0)}}\varepsilon^{(1-a)(l_{1x}+l_{1y}+l_{2x}+l_{2y})+b(n_{1x}+n_{1y}+n_{2x}+n_{2y})}\\
        &\times \Lambda_{l_{1x}l_{1y}n_{1x}n_{1y}}\Lambda_{l_{2x}l_{2y}n_{2x}n_{2y}}\widehat{\Gamma}_{l_{1x}l_{1y}n_{1x}n_{1y}}\widehat{\Gamma}_{l_{2x}l_{2y}n_{2x}n_{2y}}.
    \end{split}
\end{equation}
For the continuous limit to exist, $e^{2i\alpha^{(0)}}\Lambda_{0000}^2\widehat{\Gamma}_{0000}^2$ must equal identity, so we have the following constraint:
\begin{equation}\label{eq:ctcs_angle_constraints}
    \begin{split}
        &e^{2i\alpha^{(0)}}\Lambda_{0000}^2\widehat{\Gamma}_{0000}^2\\
        &=e^{2i\alpha^{(0)}}R_z(\omega_{x}^{(0)})R_y(\nu_{x}^{(0)})R_z(\phi_{x}^{(0)})R_z(\omega_{y}^{(0)})R_y(\nu_{y}^{(0)})R_z(\phi_{y}^{(0)})\\
        &\times R_z(\omega_{x}^{(0)})R_y(\nu_{x}^{(0)})R_z(\phi_{x}^{(0)})R_z(\omega_{y}^{(0)})R_y(\nu_{y}^{(0)})R_z(\phi_{y}^{(0)})=\mathds{1}
    \end{split}
\end{equation}
This constraint yields a similar constraint equations as from the continuous time limit in lemma~\ref{lma:constraints}, it is the following:
\begin{equation}\label{eq:constraint_function_simple}
    \begin{split}
        f&=\cos(\frac{\nu_{x}^{(0)}}{2})\cos(\frac{\nu_{y}^{(0)}}{2})\cos(\frac{\phi_{x}^{(0)}+\phi_{y}^{(0)}+\omega_{x}^{(0)}+\omega_{y}^{(0)}}{2})\\
        &-\sin(\frac{\nu_{x}^{(0)}}{2})\sin(\frac{\nu_{y}^{(0)}}{2})\cos(\frac{\phi_{y}^{(0)}-\phi_{x}^{(0)}+\omega_{x}^{(0)}-\omega_{y}^{(0)}}{2})\\
        &-\cos(\pi l-\alpha^{(0)})=0.        
    \end{split}
\end{equation}

See that the same constraints obtained in Lemma \ref{lma:constraints} for the continuous time limit are solutions of the above equation. Let us then choose $\alpha^{(0)}$ such that $\cos(\pi l-\alpha^{(0)})=0$, $\nu_x^{(0)}  =2\pi m+\lambda \pi$ and $\nu_{y}^{(0)}=2\pi t+(1-\lambda)\pi$, with $\lambda \in \mathbb{B}$. Thus the matrices $\widehat{\Gamma}_{l_xl_yn_xn_y}$ become:
\begin{equation}
    \begin{split}
        \widehat{\Gamma}_{l_xl_yn_xn_y}=i(-1)^{t+m+1}R_z(\omega_x^{(0)})\sigma_z^{l_{1x}}\sigma_y^{n_{1x}}R_z(\phi_x^{(0)}+\omega_y^{(0)})\sigma_z^{l_{1y}}\sigma_y^{n_{1y}+1}R_z(\phi_y^{(0)}).
    \end{split}
\end{equation}
Now we use this to obtain the following form of $\widehat{\Gamma}_{l_{1x}l_{1y}n_{1x}n_{1y}}\widehat{\Gamma}_{l_{2x}l_{2y}n_{2x}n_{2y}}$:
\begin{equation}
    \begin{split}
        \widehat{\Gamma}_{l_{1x}l_{1y}n_{1x}n_{1y}}\widehat{\Gamma}_{l_{2x}l_{2y}n_{2x}n_{2y}}&=\sigma_z^{l_{1x}+l_{1y}+l_{2x}+l_{2y}}\sigma_y^{n_{1x}+n_{1y}+n_{2x}+n_{2y}}\\
        &\times R_z[(-1)^{n_{1x}+n_{1y}+n_{2x}+n_{2y}}\omega^{(0)}_x\\
        &+(-1)^{n_{1y}+n_{2x}+n_{2y}}(\phi_x^{(0)}+\omega_y^{(0)})\\
        &+(-1)^{n_{2x}+n_{2y}+1}(\phi_y^{(0)}+\omega_x^{(0)})\\
        &+(-1)^{n_{2y}+1}(\phi^{(0)}_x+\omega_y^{(0)})].
    \end{split}
\end{equation}
%\MM{It does not appear that this constraint equation reveals any intrinsic symmetry to the $\widehat{\Gamma}_{l_{1x}l_{1y}n_{1x}n_{1y}}\widehat{\Gamma}_{l_{2x}l_{2y}n_{2x}n_{2y}}$ term that would show cancellations.}
%Now we implement the constraint into Eq.~\ref{eq:unitaryCoeffs}. Let $\nu_{x}^{(0)}=2\pi m+\Lambda \pi$ and $\nu_{y}^{(0)}=2\pi t+(1-\Lambda)\pi$. Plugging in this constraint into Eq.~\ref{eq:unitaryCoeffs}, we obtain the following:
%\begin{equation}
%    \begin{split}
%        \widehat{\Gamma}_{l_xl_yn_xn_y}&=\sigma_z^{l_x+l_y}\sigma_y^{n_x+n_y}R_z[(-1)^{n_x+n_y}\omega_{x}^{(0)}]R_y[(-1)^{l_y}\nu_{x}^{(0)}]\\
%        &\times R_z[(-1)^{n_y}(\phi_{x}^{(0)}+\omega_{y}^{(0)})]R_y[\nu_{y}^{(0)}]R_z[\phi_{y}^{(0)}]\\
%        &=\sigma_z^{l_x+l_y}\sigma_y^{n_x+n_y}R_z[(-1)^{n_x+n_y}\omega_{x}^{(0)}]R_y[(-1)^{l_y}\nu_{x}^{(0)}]\\
%        &\times R_z[(-1)^{n_y}(\phi_{x}^{(0)}+\omega_{y}^{(0)})]R_y[\nu_{y}^{(0)}]R_z[\phi_{y}^{(0)}]\\
%    \end{split}
%\end{equation}

Let us now discuss how choices of $a$ and $b$ change Eq.~\ref{eq:(sxcxsycy)^2}. The only terms in Eq.~\ref{eq:(sxcxsycy)^2} that will contribute to the continuum limit will be those of order $\varepsilon$, which yields a constraint concerning which terms will be non-zero after the continuum limit is taken, given a choice of $a$ and $b$:
\begin{equation}\label{eq:ab_constraint}
    (1-a)(l_{1x}+l_{1y}+l_{2x}+l_{2y})+b(n_{1x}+n_{1y}+n_{2x}+n_{2y})=1.
\end{equation}
Also, since $l_{vm},n_{vm}\in\mathbb{Z}_+$ (where $v=1,2$ and $m=x,y$), $(1-a)$ and $b$ must be in $\mathbb{Q}$ as well for this equation to hold. Upon further analysis, we see that $l_{vm}$ produces a spatial derivative with respect to $m$ in the term, so cross terms with multiple derivatives will be terms in the sum with multiple non-zero $l_{vm}$'s. $n_{vm}$ determines the presence of the driving parameter $\theta_{m}^{(1)}$ in the term. Again, as for the continuous limit, remark the existence of the cross terms derivatives which cancel for an opportune choice of $a$ and $b$. 
 
Now coming back to real space, with an inverse Fourier transform and taking the limit $\to 0$, the PDE recovered in continuous spacetime is the following:

\begin{equation}\label{eq:csct_ham_final}
    \begin{split}
        i \partial_t \Psi(x,y,t)&=i \frac{1}{2}\sum_{\substack{l_{1x},l_{1y}\\l_{2x},l_{2y}\\\neq (0,0,0,0)}}\sum_{\substack{n_{1x},n_{1y}\\n_{2x},n_{2y}\\\neq (0,0,0,0)}}\delta_{(1-a)(l_{1x}+l_{1y}+l_{2x}+l_{2y})+b(n_{1x}+n_{1y}+n_{2x}+n_{2y}),1}\\
        &\times \Lambda'_{l_{1x}l_{1y}n_{1x}n_{1y}}\Lambda'_{l_{2x}l_{2y}n_{2x}n_{2y}}\widehat{\Gamma}_{l_{1x}l_{1y}n_{1x}n_{1y}}\widehat{\Gamma}_{l_{2x}l_{2y}n_{2x}n_{2y}}\Psi(x,y,t)
    \end{split}
\end{equation}
where
\begin{equation}
    \Lambda'_{l_xl_yn_xn_y}=\frac{(\partial_x)^{l_x}(\partial_y)^{l_y}(-\frac{i\nu_{x}^{(1)}}{2})^{n_x}(-\frac{i\nu_{y}^{(1)}}{2})^{n_y}}{l_x!l_y!n_x!n_y!}
\end{equation}
and $\delta_{ij}$ being the Kronecker delta.

The above equation is very powerful, as it identifies the type of PDE obtained for any possible choice of $a$ and $b$. As an example, we analyze the $a=b=\frac{1}{2}$ scenario. For this case, the only terms which will contribute are terms with two of the $l_{vm}$'s equalling 1 and the $n_{vm}$'s equalling 0 (6 terms), terms with one $l_{vm}$ equalling 1 and one $n_{vm}$ equalling 1 (16 terms), and two of the $n_{vm}$'s equalling 1 and the $l_{vm}$'s equalling 0 (6 terms).

When analyzing these terms further, we see that only 1 term proportional to $(\theta^{(1)}_{x})^2$ and one term proportional to $(\theta^{(1)}_{y})^2$ will be present in the sum. These terms have no derivatives in them. There is only one of these terms in the sum, so there is no way to cancel them out for all choices of $\nu$, $\omega$, and $\phi$. The matrices present in these terms are also products of unitary matrices, which cannot equal zero, therefore these terms will be present in the ensuing PDEs for all choices of constraints and general choice of parameters $\nu$, $\omega$, and $\phi$. In a similar note, when analyzing the terms with two $l_{vm}$'s equalling 1, we see that there is only 1 term proportional to $\partial_x^2$ and 1 term proportional to $\partial_y^2$. Under the same line of reasoning as before, we can conclude that these terms will also be present for all choices of constraints and general choice of parameters $\nu$, $\omega$, and $\phi$. 

After some more analysis, we see that the cross derivative terms (terms proportional to $\partial_x\partial_y$) will not be present if the constraints $\theta^{(0)}_x=2\pi m$,$\theta^{(0)}_y=2\pi t+\pi$ or $\theta^{(0)}_x=2\pi t+\pi$, $\theta^{(0)}_y=2\pi m$ are used, which is not true if any other constraints are used. It can also be shown that if the constraints $\theta^{(0)}_x=2\pi m$ and $\theta^{(0)}_y=2\pi t+\pi$ are used, the ensuing continuum limit PDE will be the form:
\begin{equation}
\partial_t\Psi(x,y,t)=(P_1\partial_x+P_2\partial_y+P_3\partial_x^2+P_4\partial_y^2+MP_5)\Psi(x,y,t).
\end{equation}

The above equation shows that with an opportune choice of the zero order constraints and the parameters $l_{vm}$ we can recover not only the advection term $\partial_i \Psi$, but also dispersive terms $\partial^2_i \Psi$ and a mass term. 

If $l_{vm}\in \{0,1\} $ and $n_{vm}\in \{0,1\} $, where $l_{vx}$ and $l_{vy}$ are not equal $1$ in the same term of the sum we will recover a full propagative equation, namely the Dirac equation of the kind 
\begin{equation}
\partial_t \Psi  = \left(P_x(\theta)\partial_x +P_y(\theta)\partial_y\right)\Psi
\end{equation}
where all the different $P_i$ are non commutative hermitian matrices.

Finally, we resuming the above Lemmas proven in this section with the following Theorem: 

\begin{Th}[Strong Plastic Quantum Walks in 2D]
If $\tau=2$, $d=2$ and all the following conditions are verified
\begin{enumerate}
\item $\nu_x^{(0)} = 2\pi m+\lambda \pi$
\item $\nu_{y}^{(0)}=2\pi t+(1-\lambda)\pi$
\item $\alpha^{(0)}$ such that $\cos(\pi l-\alpha^{(0)})=0$, $\nu_x^{(0)}  =2\pi m+\lambda \pi$ and $\nu_{y}^{(0)}=2\pi t+(1-\lambda)\pi$, with $\lambda \in \mathbb{B}$
\end{enumerate}
with $\lambda \in \mathbb{B}$, then the QW displays strong plasticity.
\end{Th}

\section{Discussion and perspectives}

In this chapter we have demonstrated the existence of a plastic QW on the grid. The equations that QW can simulate are much more general than those in 1D and include second degree terms in the spatial derivatives. This result indicates that, contrary to what has been believed until now, QWs can also simulate dispersive PDEs, a very large class of equations that includes the Shr\"odinger equation. In particular, the more general class of PDEs that can be simulated include the advection-dispersion equations.  This result is totally unexpected and novel. The possibility to simulate second degree terms suggests the we have a way to quantum simulate even irreversible systems, by using microscopic degrees of freedom as a reservoir. Can a sequence of appropriate unitaries approximate a non-unitary dynamic? This question opens new perspectives and applications to all those algorithms and computational schemes, such as the Monte Carlo simulations, which today are essentially still relying on classical Random Walks. Such Plastic QWs, employing superpositions of states, might hopefully provide them a quadratic improvements in computational time, while preserving the dispersive feature of the classical algorithms. 

\medskip

\chapter{Quantum Walking over triangles}\label{chap:triangle}

%\toquotebig{citation}{auth}

\toabstract{
Some QWs admit a continuum limit, leading to well-known partial differential equations in Physics, such as the Dirac equation. We show that these simulation results need not rely on the grid: the Dirac equation in (2+1)--dimensions can also be simulated, through local unitaries, on the triangular lattice. This result opens the door for a generalization of the Dirac equation to arbitrary discrete surfaces.
}

\clearpage

In this chapter we depart from the grid and we will introduce a new family of QW, which relies on triangulations. The motivation is to model discrete transport in all sorts of topologies as simplicial complexes. We must consider this chapter as an introduction to the next one. In fact, if we have previously proved that different QWs can simulate a large family of hyperbolic differential equations, the ultimate goal is to demonstrate that QWs can be a powerful tool to simulate transport equations on any kind of surface. Let us imagine for example a sphere and let ask ourselves how to simulate on it the transport of a quantum system with two internal states, e.g. a qubit. The most natural way is to triangulate the curved surface and implement a QW in discrete time on that surface. A necessary step to gain this goal is to define a new family of QWs capable of propagating on a surface tessellated by regular triangles, e.g. equilateral triangles. Studying QWs on triangular surfaces is not only interesting to simulate hyperbolic PDEs on arbitrary manifolds. The interest is of a very large scientific community ranging from the use of new carbon-based materials and their topological properties, to algorithmics and some quantum gravity theories. More and more new technologies make use of materials such as graphene or fullerene or even compounds with crystalline structures such as triexhagonal tessalation, namely Kagome lattice. In each of these structures, the simplest tile is triangular in shape. To ask how quantum information can be transmitted on such topologies is one of the most immediate applications of appropriate QW defined on triangular structures. There have been researches in this direction and in many of them it has been understood how such QWs are able to capture the topological properties of such materials, allowing also a classification. In all previous models of QWs on triangles, the most serious problem was to use three-state systems, which from a technological point of view made their implementation very complicated and costly. In this chapter we will show how, despite these structures are spanned by three base state vectors, defining the three possible directions of displacement, a QW with a physical system with only two internal states, such as a qubit, is possible and can be defined. The idea is similar to the one that led to the introduction of the AQW, but to solve this problem we will have to modify the definition of the walker itself and his configuration space. 

In the first part of the chapter we will introduce the model. Then, we will prove that such a QW admits as a continuous limit a homogeneous hyperbolic PDE, which suggests that such a result does not need necessary a square grid. We will focus in particular on a class of QW on triangles, namely Dirac QWs, which converge on a hyperbolic PDE called the Dirac Equation, similar to what we did in the previous chapter.  

\section{The model}\label{subsec:triangles}

\begin{figure}
\center
\includegraphics[width=0.5\columnwidth]{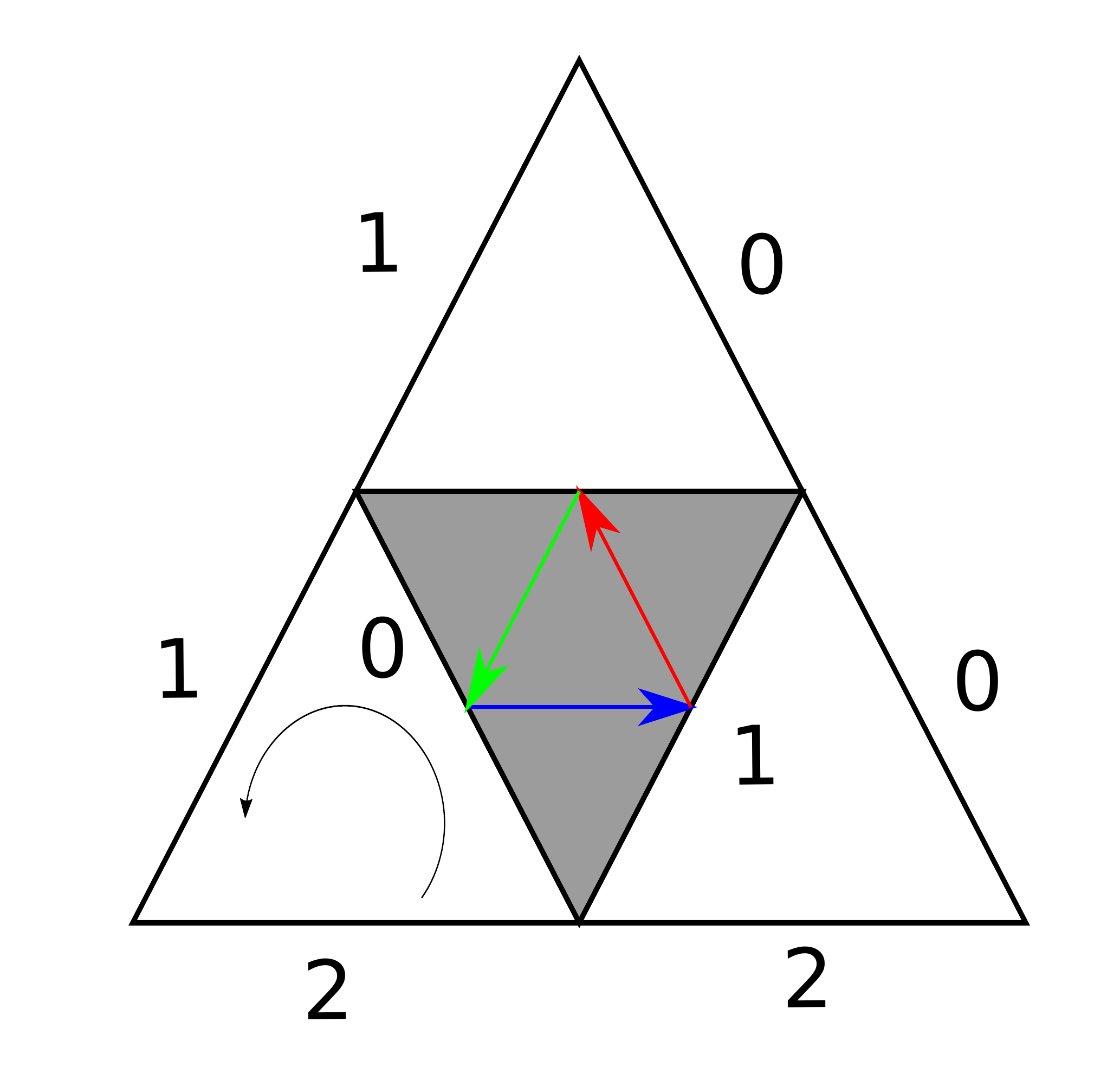}
\caption{ The Triangular QW. Starting at the edge $k=0$.
The circle line represents the counter-clockwise rotation operator. }
\label{Fig:Honeyandtriangles} 
\end{figure}

Our triangles are equilateral with sides $k=0,1,2$, see Fig. \ref{Fig:Honeyandtriangles}. Albeit the drawing shows white and gray triangles, these differ only by the way in which they were laid --- they have the same orientation for instance. The QW Hilbert space is $\mathcal{H} = \mathcal{H}_e \otimes \mathcal{H}_2$ where $\mathcal{H}_e$ is spanned by the basis states $\ket{e}$ with $e$ an edge of the grid and $\mathcal{H}_2$ is spanned by the usual coin basis states $\ket{0}$ and $\ket{1}$. Differently from the QW on the grid, here the QW lies on the edge of the simplex. Notice that each triangle hosts a $\mathbb{C}^{3}$ vector, e.g. $(\psi^{0,v}_{j,k})_{k=0\ldots2}$ and $(\psi^{1,v}_{j,k})_{k=0\ldots2}$. For two triangles $v$ and $v'$ sharing an edge, this edge contains twice the information (one for each triangle). This issue is addressed by associating each coin state to a specific triangle. Let us label each triangle with ($0$ or $1$) such that any two adjacent triangles have different internal state. Moreover, if $v$ is a triangle and $k \in \{1,2,3\} \cong \mathbb{Z}/3\mathbb{Z}$, the generic state $\ket{e} \equiv \ket{k_v}$ is uniquely defined by the number $v$ and the number $k$. The generic state of the walker at a given time $j'$ reads:
\begin{equation}
\Psi_{j'} = \sum_{v,k} \left(\psi^{0,v}_{j',k} \ket{0,k_v} + \psi^{1,v}_{j',k} \ket{1,k_v}\right).
\end{equation}
The dynamics of the Triangular QW is the composition of two operators. The first operator, $R$, simply rotates every triangle anti-clockwise. Phrased in terms of the hosted $\mathbb{C}^{3}$ vectors, the component at side $k$ hops to side $(k+1~\text{mod}~3)$, as shown in Fig. \ref{Fig:Honeyandtriangles}. Equivalently each rotation may be seen as the synchronous application of the translation operator $S_{u_i}$ defined as follows:
\begin{equation}
\widehat S_{u_k}\left(\begin{array}{c}
\psi^{0,v}_{j,k}\\
\psi^{1,v}_{j,k}
\end{array}\right)=\left(\begin{array}{c}
\psi^{0,v}_{j,k+1\text{mod}3}\\
\psi^{1,e(v,k)}_{j, k+1\text{mod}3}
\end{array}\right)\label{eq:triangulatQW}
\end{equation}
where $e(v,k)$ represents the neighbor along the $k^{th}$-edge of the triangle $v$. Notice that we can rely the displacement on the triangular grid on the vector basis $u_{x}$ and $u_{y}$ spanning the square Euclidean grid. Let us introduce the position $r\equiv r(x,y)$, which corresponds to the mean value of each edge $k$ of a triangle $v$, then the two components wave fonction $(\Psi^{v}_{j,k}) = \Psi_{j,k}(r)$ represent the vector at position $r$, on the edge $k$ at time $j$. Notice that although $r$ characterizes completely the position of the walker, the shift is still $k$-dependent and will always need both informations. Finally the displacement operator will read: 
\begin{equation}
\tilde S_{u_k}\left(\begin{array}{c}
\psi^{0}_{j,k}(r)\\
\psi^{1}_{j,k}(r)
\end{array}\right)=\left(\begin{array}{c}
\psi^{0}_{j,k+1\text{mod}3}(r-\Delta \hat u_k)\\
\psi^{1}_{j, k+1\text{mod}3}(r+\Delta \hat u_k)
\end{array}\right)\label{eq:shiftXY}
\end{equation}
where $\Delta$ is the lattice discretisation step of the triangle grid. 
Now the basis vector $u_{i},\,\,\,i=0,1,2$ are given by 
\begin{equation}\label{eq:change}
u_{i}=\cos(i\frac{2\pi}{3})u_{x}+\sin(i\frac{2\pi}{3})u_{y} = R^s u_s,
\end{equation}
where $s = x, y$ and $R$ is the coordinates change matrix, and finally the shift operators can be written as a convex combination of the usual displacement $\widehat S_{u_x}$ and $\widehat S_{u_y}$ we introduce in Chap. \ref{chap:quantumwalks} :  
\begin{equation}\label{eq:shiftXY}
\tilde S_{u_k} =\cos(i\frac{2\pi}{3})\widehat S_{u_x} +\sin(i\frac{2\pi}{3}) \widehat S_{u_y}.\nonumber 
\end{equation}

The second operator is the application of the usual $2\times2$ quantum coin $\hat C$ over each two-components vector state lying on every edge shared by two neighboring triangles, but now the coin will be in general different for each edge $k$. Altogether, the Triangular QW dynamics is given by the following recursive relations: 
\begin{equation}
\Psi_{j+1,k}(r) = \Pi_{i=0,1,2} \tilde S_{u_k + i\text{mod}3} C_{k + i\text{mod}3} \Psi_{j,k}(r).
\label{eq:triangulatQW}
\end{equation}

The above equation is a very general one. In the next corollary we are going to show that there exists a sub set of quantum coins $C$, which does not depend on the edge $k$ and for which the Eq \ref{eq:triangulatQW} admit a continuous limit.

\begin{Cor}[Continuous Limit of a Triangular QW]\label{cor:CLtriangle}
Let consider the Eq. \ref{eq:triangulatQW} driving an Alternate Quantum Walk over a triangular lattice. Admit that the quantum coins $\hat C$ doesn't depend on its position. If $\theta_{0,u_1} =\theta_{0,u_2}=\theta_{0,u_3} = (1-s)\pi + 2q\pi$, $\xi_{0,u_1} =\xi_{0,u_2}=\xi_{0,u_3} = \frac{\pi}{3} + q'\pi$ and $\alpha_{0,u_1} =\alpha_{0,u_2}=\alpha_{0,u_3} = \frac{\pi}{3} + s\pi + 2q"\pi$, where $s\in\mathbb{B}$ and $\{q,q',q"\}\in\mathbb{Z}$, then the Eq. \ref{eq:triangulatQW} admits as continuous limit the following PDE:
\begin{equation}\label{eq:falseDE}
\partial_t \Psi = P( \partial_x + \partial_y)\Psi \equiv P \nabla \Psi
\end{equation}
where $P$ is a Hermitian and traceless matrix.
\end{Cor}
\textbf{Proof.}
By hypothesis $C$ doesn't depend on the walker position, then $C^{(0)}$ is also homogeneous. If we introduce in the Eq. \ref{eq:triangulatQW} the discretization parameters $\Delta = \varepsilon$, for the spatial dimensions and $\Delta_t = \varepsilon$ for the time dimension, and we develop the same equation around $\varepsilon$ then we get up to the first order : 
\begin{equation}
\begin{split}
\Psi_{k}(t,r) &+ \varepsilon \partial_t\Psi_{k}(t,r) = (C^{(0)})^3\Psi_{k}(t,r) +  \varepsilon ( (C^{(0)})^2\sigma_zC^{(0)}\partial_{u_0}\Psi_{k}(t,r) \\
+& C^{(0)}\sigma_z(C^{(0)})^2\partial_{u_1}\Psi_{k}(t,r) +(C^{(0)})^3\sigma_z\partial_{u_2} \Psi_{k}(t,r))
\label{eq:triangulatQW}
\end{split}
\end{equation}
Then in order to consider the formal limit $\varepsilon \to 0$, we need to satisfy:
\begin{equation}
(C^{(0)})^3 = \mathds{1}
\end{equation}
from which we derive the following conditions
\begin{equation}
\begin{split}
\cos(2\theta_0) + 2\cos^2(\theta_0)\cos(\xi_0) = 0\\
e^{3 i \alpha -i \xi } \cos (\theta ) \left(e^{4 i \xi } \cos ^2(\theta )-\left(1+2 e^{2 i \xi }\right) \sin ^2(\theta )\right)=1
\end{split}
\end{equation}
The above equations are satisfied if $\theta_{0} = (1-s)\pi + 2q\pi$, $\xi_{0} = \frac{\pi}{3} + q'\pi$ and $\alpha_{0} = \frac{\pi}{3} + s\pi + 2q"\pi$. For theses zero$^{th}$ conditions 
\begin{equation}\label{eq:zeroTriangle}
C^{(0)} = e^{i\frac{\pi}{3}}R_{z}(\frac{2\pi}{3}).
\end{equation}
Then the Eqs. \ref{eq:triangulatQW} read:
\begin{equation} 
\partial_t\Psi_{k}(t,r) = (\sigma_z\partial_{u_0} + \sigma_z\partial_{u_1} +\sigma_z\partial_{u_2} )\Psi_{k}(t,r) 
\label{eq:triangulatQW2}
\end{equation}
where we used the fact that $(C^{(0)})^2\sigma_zC^{(0)}=C^{(0)}\sigma_z(C^{(0)})^2 = \sigma_z$.

Notice that the PDE obtained in the Corollary \ref{cor:CLtriangle}, although defined in 2D is effectively in one spatial dimension. In fact, it is sufficient to define a new variable $v = x + y$ to reduce the Eq \ref{eq:falseDE} to the one dimensional DE obtained in \ref{eq:theo2steps}. In order to extend the DE to a proper 2D transport we need a different non commuting Hermitian matrix in front to each spatial derivative, e.g. $\sigma_x$ and $\sigma_y$.
Using the \ref{cor:CLtriangle} we can prove that this is possible by mean of a basis change on the coin states.

\begin{Th}[Dirac Quantum Walk over triangulation]\label{Th:DiracTrQW}
Consider the AQW described in \ref{eq:triangulatQW} where 
\begin{equation}\label{eq:zeroTriangle}
C^{(0)} = \begin{pmatrix}
e^{i\frac{\pi}{3}}R_{z}(\frac{2\pi}{3})
\end{pmatrix}
\end{equation}
If $\tilde \Psi_{k}(t,r) = U\Psi_{k}(t,r)$, $C^{'(0)} = UC^{(0)}U^\dagger$ and $U = e^{-i \alpha\sigma_y/2} C^\dagger$ where $ \alpha = - \arccos(\sqrt{5}/3)$, then the AQW \ref{eq:triangulatQW} reads as:
\begin{equation}
\tilde \Psi_{j+1,k}(r) = \Pi_{i=0,1,2} \tilde S_{u_k + i\text{mod}3} UC  U^\dagger \tilde \Psi_{j,k}(r).
\end{equation}
and it admits as continuous limit in spacetime the following PDE:
\begin{equation}
\partial_t\Psi(t,x,y) =(\sigma_x \partial_x + \sigma_y \partial_y)\Psi(t,x,y)
\end{equation}
\end{Th}
\textbf{Proof.}
Using the Corollary \ref{cor:CLtriangle} we know that \ref{eq:triangulatQW} admits a continuous limit iff the zero$^{th}$ order of the coin is given by the Eq. \ref{eq:zeroTriangle}. Notice that a generic basis change applied to the coin states $\{\ket{0},\ket{1}\}$ and implying that $C^{'(0)} = UC^{(0)}U^\dagger$ and $\tilde \Psi_{k}(t,r) = U\Psi_{k}(t,r)$ don't change the zero$^{th}$ order conditions, in fact:
\begin{equation}
\Psi_{k}(t,r) = U^\dagger (C^{'(0)})^3 U \Psi_{k}(t,r) 
\end{equation}
is satisfied because $U^\dagger U = \mathds{1}$. \\
Let us now Taylor expand up to the first order the equation \ref{eq:triangulatQW} in the new basis:
\begin{equation}\label{eq:finalTriangle}
\partial_t  \Psi_{k}(t,r) =\left(\tau_2\partial_{u_2} + \tau_1\partial_{u_1}  + \tau_0 \partial_{u_0} \right)\Psi_{k}(t,r) +o(\varepsilon)
\end{equation}
where 
\begin{equation}\label{eq:tauU}
\begin{split}
\tau_0 =  (C^{(0)} )^2  U^\dagger\sigma_z U  C^{(0)} \\
\tau_1 = C^{(0)} U^\dagger\sigma_z U (C^{(0)} )^2\\
\tau_2 = U^\dagger\sigma_z U
\end{split}
\end{equation}
To prove the theorem we need :
\begin{equation}\label{eq:condshift}
\begin{split}
\sum_{i=0,1,2} \tau_i \partial_{u_i} =  (\sigma_x \partial_x + \sigma_y \partial_y).
\end{split}
\end{equation}
Notice that, using the Eq. \ref{eq:change} in the above equation, we can derive a relation between the $\tau_i$ and the $\sigma^s$:
\begin{equation}\label{eq:sigmatau}
R^s_i \tau^{i} = \sigma^s.
\end{equation}
%where we use the Corollary \ref{cor:CLtriangle} in the last equation.\\
To solve the system \ref{eq:condshift}, we first translate the $(\partial_{u_i})_{i=0,1,2}$ in terms of the coordinates $(x,y)$, using the Eq. \ref{eq:shiftXY}
\begin{equation}
\partial_{u_k} =\cos(i\frac{2\pi}{3})\partial_{x} +\sin(i\frac{2\pi}{3}) \partial_{y}.\nonumber 
\end{equation}
and then the conditions \ref{eq:condshift} lead to unique $(\tau_{i})$
matrices, up to a sign: 
\begin{equation}\label{eq:tau}
\begin{split}
\tau_{0} & =\frac{2}{3}\sigma_{x}+\kappa\sigma_{z} \\
\tau_{1} & =-\frac{1}{3}\sigma_{x}+\frac{\sqrt{3}}{3}\sigma_{y}+\kappa\sigma_{z}\\
\tau_{2} & =-\frac{1}{3}\sigma_{x}-\frac{\sqrt{3}}{3}\sigma_{y}+\kappa\sigma_{z}.
\end{split}
\end{equation}
with $\kappa=\pm\frac{\sqrt{5}}{3}$. Choose $\kappa=\frac{\sqrt{5}}{3}$,
and notice that 
\begin{equation}
\sum_{i}\tau_{i}=\frac{\sqrt{5}}{3}\sigma_{z}.\label{eq:tauisigmaz}
\end{equation}
and finally from Eq. \ref{eq:tau} and the Eq. \ref{eq:tauU}, we can derive $U = e^{-i \alpha\sigma_y/2}C^\dagger$, where $ \alpha =- \arccos(\sqrt{5}/3)$ and thus, after the formal limit $\varepsilon \to 0$, the Eq \ref{eq:finalTriangle} is the following DE in (2+1) dimensions :
\begin{equation}
\partial_t \Psi(t,x,y) =(\sigma_x \partial_x + \sigma_y \partial_y)\Psi(t,x,y).
\end{equation}

%\subsection{Kagome lattice}\label{subsec:kagome}
%
%It could be useful to connect the above family of QW with the Alternate QW introduced earlier. Let us imagine that the walker is located at the center of each edge in the triangular lattice. If we replace the center of each edge by a node and we connect the nodes along the axis of propagation of the walker, we recover a trihexagonal tiling, even known as Kagome lattice. This lattice combines regular triangles with regular hexagon and the extension to the three dimensional space, which is filling space, is formed by regular tetrahedra and truncated tetrahedra, and called a hyper-kagome lattice. Here we substantially review the above results from this new perspective. \\
%Although each node of the Kagome lattice has two neighbors, the shift operator will be not the same. In other terms we need to $3$-color the lattice and for each color we need a different shift operator for each color. The QW lies again in a composite Hilbert space $\mathcal{H} = \mathcal{H}_{\mathbb{Z}^2} \otimes \mathcal{H}_{2}$. Where $\mathcal{H}_{\mathbb{Z}^2}$ is spanned by $\ket{m,n}$ and $\mathcal{H}_{2}$ is spanned by the usual coin states $\{\ket{0},\ket{1}\}$. At each point we apply a quantum coin as defined in \ref{def:coin} and the three shift operators are the following  TBC
%%\begin{equation}
%%\hat S^{white} = \ket{}
%%\end{equation}
%%

\section{Discussion and open problems}

In this chapter we first defined a family of QWs on triangles using a two-states system, e.g. a qubit. Then we computed their continuous spacetime limit. This limit turned out to coincide with a pair of hyperbolic PDEs. These describe the transport of the walker at constant velocity over an Euclidean plane. We have therefore proved that we can simulate the Dirac equation and more generally any hyperbolic PDEs, over a grid. This result is a necessary step to extend these Dirac QW to any triangulation. There is still a long way to go, but the perspectives are clear. Starting with a dimensional extension of such QW in 3D spatial dimensions. Moreover, these results are fundamental to finally be able to talk about plasticity away from the grid. We believe that a possible union between plasticity and the intrinsic malleability of triangulations can open different doors. One of them is to build more versatile simulation models at the frontier between quantum lattice theories and differential geometry, building bridges between them. There are many questions that can be asked about this frontier: one in particular has caught our attention. How to simulate a hyperbolic PDE on curved surface? Let us imagine that we continuously deform a plane, e.g. tessellated by equilateral triangles, until it becomes \emph{curved} like a sphere. To do this, we will be forced to stretch distances, modify angles and create defects. Such a deformation is not unitary, because the coefficients of the deformation matrix are in general arbitrary. Is it possible to simulate such an operation by means of local unitaries? This question will be answered in the next chapter. 

\medskip

%\noindent We have seen how CA are not only a physics-like model of computation, but also a computer-science-like model of physical phenomena. Plus we have given an overview of the rich mathematical structure they have. To quite some extent one can view this thesis as a port / generalization of several of these results to the quantum regime. We have also hinted at a few difficulties awaiting for us along that path.

\newpage

% X pages

\chapter{Simulating transport on curved surfaces}\label{chap:manifold}

%\toquote{citation}

\toabstract{
We apply a spacetime coordinate transformation upon the triangular lattice of the previous QW, and show that it is equivalent
to introducing spacetime-dependent local unitaries \textemdash whilst
keeping the lattice fixed. By exploiting this duality between changes
in geometry, and changes in local unitaries, we show that spacetime-dependent
QWs simulate the Dirac equation in $(2+1)$\textendash dimensional
curved spacetime. Interestingly, the duality crucially relies on the
non linear-independence of the three preferred directions of the triangular lattices.
 At the practical level, this result opens the possibility to simulate free field theories on curved manifolds. }

\clearpage

In the previous chapter we have proven that some family of QWs admits as continuous limit in spacetime an equation of the form:
\begin{equation}\label{eq:flatDE}
\partial_t \Psi(t,\mathbf{x}) =  \sum_{i=1}^{d} P_i \partial_{x_i}\Psi(t,\mathbf{x})
\end{equation}
where $C$ may be seen as a constant and uniform velocity matrix for the continuous field $\psi(t,\bd{x})$. Any function of the form $\omega(\bd{x} - ct)$ is a solution of the above equation and it describes the transport of the function $\omega$ at constant speed $v$. In other words the evolution operator which drives the field is homogeneous at every point of the space. What does it happen if we consider the same equation on an arbitrary manifold? Such a PDE in an arbitrary dimension $d$ can be written as follows:
\begin{equation}\label{eq:curvedDE}
\partial_t \Psi(t,\bd{x}) = \sum_{i=1}^{d}\left( P_i(\bd{x}) \partial_{x_i}\Psi(t,\bd{x}) + \Psi(t,\bd{x}) \frac{1}{2} \partial_{x_i}P_i(\bd{x}) \right). 
\end{equation}
where each $P_i(\bd{x})$ is a Hermitian matrix, such that $|P_i(\bd{x})|\leq \mathds{1}$. 
%A rigorous derivation of this equation is done in Appendix \ref{app:A}. 
Notice that the space dependent coefficients in front of the derivatives may be seen as component of a non-homogeneous velocity tensor. This suggests that the mechanism now driving the walker needs to be non homogeneous, in particular the quantum coin parameters. Our goal in the following is to build a quantum scheme such that, given a particular PDE of the form \ref{eq:curvedDE} we wish to simulate, we are able to retro-engineer the corresponding QW. To illustrate the main idea let us start by a the simple academic case of the infinite line. 

\section{Mimicking curved transport on the grid}\label{sec:curved_grid}

Let us consider a QW over the line, e.g. that one introduced in the previous chapter. We already proved that it formally converges, in the continuous spacetime limit, to the Eq. \ref{eq:strong} which is of the form $\ref{eq:flatDE}$, where the velocity corresponds to the real constant $\bar \theta$. Our conjecture is that, to simulate a non-homogenous velocity field, we would wish to make $\bar \theta$ dependent on each point of the spacetime grid:
\begin{equation}\label{eq:theta}
\theta(t,x) = \theta_0 + \varepsilon^b \bar \theta(t,x). 
\end{equation}
Then coming back to the proof of Theorem \ref{th:CL1D_2steps} for $a=1$ we can see that the coefficients $ \varepsilon \partial_x [\{\sigma_z C^{(0)}, C^{(1)}(t,x)\}+\{\sigma_z C^{(1)}(t,x), C^{(0)}\}]$ leads now to a spatial derivative of $C^{(1)}(t,x)$, due to the inhomogeneity introduced in the jet \ref{eq:theta}. Computing the formal limit for $\varepsilon \to 0$, leads to recover the following couple of inhomogeneous PDEs:
\begin{equation}\label{eq:strongcurved}
\begin{split}
\partial_t \psi^0 &= 2 i e^{i \zeta}( \bar \theta(t,x) \partial_x \psi^{1} + \psi^{1}\frac{1}{2}\partial_x  \bar \theta(t,x))\\
\partial_t \psi^1 &= -2 i e^{-i \zeta} (\bar \theta(t,x) \partial_x \psi^{0} + \psi^{0}\frac{1}{2}\partial_x  \bar \theta(t,x))
\end{split}
\end{equation}
The above equations are of the required form in \ref{eq:curvedDE}. Indeed, fixed by hand a specific metrics leads to define the $\bar \theta(t,x)$ and thus, to retro-engineer the corresponding QW just encoding the metric coefficients into the definition of the quantum coin. In order to show how it works, we propose the following example:

\begin{Ex}[Quantum Walking in and around a black hole]

The Eq. \ref{eq:curvedDE} includes the massless Dirac equation (DE) coupled to a $(d+1)$-dimensional spacetime. In $(1+1)$-spacetime and synchronous frame, the massless DE can be written as follows: 
\begin{equation}
i\partial_{t}\chi+\frac{i}{2}\{B^{x},\partial_{x}\}\chi=0,\label{eq:Oliveira}
\end{equation}
and where \textbf{$B^{x}=\gamma^{1}{e^{x}}_{1}$}. In particular, one can make the choice $\gamma^{1}=\sigma_{x}$. Here we are not discussing the physics behind this equations. We are only interested in formally simulating them on the lattice by means of local unitaries. \\Let us fix in \ref{eq:strongcurved} $\zeta$ to $-\pi/2$ so that it coincides with Eq. \ref{eq:Oliveira} and by identification we can set $e^{x}_{1} = \bar \theta(t,x)$. Now we assume that the spacetime fonction $B^{x}$, which defines the spacetime metrics coefficients, is externally fixed by hand, \ie the corresponding curvature is external and not dynamical. As exemple, we choose the spherically symmetric solution of Einstein equation in vacuo, a Schwarschild black hole. In a special coordinates frame, called Lema\^itre coordinates, the Eq. \ref{eq:Oliveira} reads:
\begin{equation}
i\partial_{t}\chi-\frac{i}{2}\{ \frac{r_g}{r} \sigma_{x},\partial_{x}\}\chi=0,\label{eq:Oliveira2}
\end{equation}
where $r (t, x) = \left[\frac{3}{2}\left(x- t\right)\right]^{2/3}r_g^{1/3}$, $r_g$ is constant and called the gravitational radius. The event horizon is located at $r = r_g$ {\sl i.e.} $x = t + (2/3) r_g$, and the singularity is located at $r = 0$ {\sl i.e.} $x = t$. The exterior of the black hole is the domain $r > r_g$ and finally the trajectory of the quantum particle, \ie the geodesics the particle would follow in this metrics, are analytically computed by $d t = \pm \left( r_g/r(t, x)\right)^{1/2} d\rho$. \\

Now, again, by identification 
\begin{equation}
\bar \theta (t, x) = \sqrt{\frac{r(r, x)}{r_g}}
\label{eq:deftheta}
\end{equation}
and thus we can prepare our local unitaire $W$, in particular the quantum coin $C$ to simulate the transport of a quantum particle in and around the radius of Schwarschild black hole. 
%The simulations as clearly shown in Fig XX, displays that the QW closely follows the geodesics of the metrics. 
\end{Ex}

\section{A Quantum Walk over a generic 2D spatial triangulation}\label{sec:TQWcurved}

Now let us depart from the grid and let us consider a QW over an equilateral triangles as introduced in Theorem \ref{Th:DiracTrQW}. Due to the high malleability of triangles to pave any curved surface  in two spatial dimensions, we may argue that triangles could open the doors to an other way to simulate a transport equation with inhomogeneous velocity as in \ref{eq:curvedDE}: for instance we can start to stretch the distances between the point of the grid in an inhomogeneous way, deforming continuously the surface over which the QW propagate. Such a deformation will introduce locally a metrics which is not simply recoverable by a change of coordinates. However this operation, which is already known as a standard technique to triangulated curved manifold, is not unitary and then cannot be accounted in our quantum numerical scheme. However we may be luckier than that and find out that such deformation may be absorbed by local non homogeneous unitary as we did on the grid. We will see how in the following, but first we will explore the idea introducing a global homogeneous deformation of the lattice. Immagine that such transformation $\Lambda$ exists, the basis vector $u_{i}$, introduced in \ref{subsec:triangles}, which span the Euclidean plane will transform as follows:
\begin{equation}
u'_{i}=\left(\begin{array}{cc}
\lambda_{11} & \lambda_{12}\\
\lambda_{21} & \lambda_{22}
\end{array}\right)u_{i}\equiv\Lambda u_{i}.\label{eq:transu}
\end{equation}
where $\lambda_{ij}$ are position independent, although they are eventually allowed to depend on time. The derivative $\partial_{u_i}$ will also transform as 
\begin{equation}
\partial_{u'_i}\equiv\Lambda \partial_{u_{i}}
\end{equation}
and then using the Theorem \ref{Th:DiracTrQW} and plugging them into Eq. \ref{eq:finalTriangle}, we arrive to the following equation:

\begin{equation}
i\partial_{t}\psi=\left[\left(\lambda_{11}\sigma^{x}+\lambda_{12}\sigma^{y}\right)\partial_{x}+\left(\lambda_{21}\sigma^{x}+\lambda_{22}\sigma^{y}\right)\partial_{y}\right]\psi,
\label{eq:diracmod}
\end{equation}

which describes the Dirac equation on a flat geometry, in fact the velocity still remain homogeneous in space time and comparing with Eq. (\ref{eq:Oliveira}) gives 

\begin{align}
B^{x} & =\lambda_{11}\sigma^{x}+\lambda_{12}\sigma^{y}\label{eq:Bsfromsigmasflat1}\\
B^{y} & =\lambda_{21}\sigma^{x}+\lambda_{22}\sigma^{y}.\label{eq:Bsfromsigmasflat2}
\end{align}
or $B^s = \Lambda^s_i \sigma^{i}$ and using the equation \ref{eq:sigmatau}
\begin{equation}\label{eq:BLambdaTau}
B^s = \Lambda^s_i R^{i}_j\tau^j.
%B^{s}=\Lambda_{k}^{s}u_{i}^{k}\tau^{i}.
\end{equation}

Notice that the homogeneous transformation $ \Lambda$ may also help us to redefine the matrices $\tau_i$. In fact from Eq \ref{eq:finalTriangle} we get:
\begin{equation}\label{eq:finalTriangle2}
\partial_t  \Psi_{k}(t,r) =\left(\tau'_2\partial_{u_2} + \tau'_1\partial_{u_1}  + \tau'_0 \partial_{u_0} \right)\Psi_{k}(t,r) 
\end{equation}
where 
\begin{equation}
\tau'_i = \Lambda \tau_i .
\end{equation}

It is now clear that in order to simulate an inhomogeneous velocity field we need to choose a space-time dependent $\Lambda(t,x,y)$ transformation. Because such a transformation is not unitary, from Equation \ref{eq:BLambdaTau}, we will explore an equivalent way to retro-engineer the deformation playing with our local unitaries.

Instead of introducing a distortion $\Lambda(t,x,y)$ on the lattice
via the modification of the $u_{i}$ vectors, the unitary matrices
$\tau^{i}$ can be transformed to produce the same effect. In other
words, we seek for a set of matrices $\beta^{i}(t,x,y)$ that fulfill
the following conditions: 
\begin{itemize}
\item (C1) We impose that 
\begin{equation}
\Lambda_{k}^{j}(t,x,y)R_{i}^{k}\tau^{i}=R_{i}^{j}\beta^{i}(t,x,y).\label{eq:sumpi}
\end{equation}
\item (C2) Each of them has $\{-1,1\}$ as eigenvalues, i.e. at any time
step and at any point $(x,y)$ of the lattice there exist three unitaries
$U_{i}(t,x,y)$ such that 
\begin{equation}
\beta^{i}(t,x,y)=U_{i}^{\dagger}(t,x,y)\sigma^{z}U_{i}(t,x,y).\label{ref:beta}
\end{equation}
\end{itemize}
Notice that condition (C1) implies that the coordinate transformation
dictated by $\Lambda_{k}^{j}(t,x,y)$ is transferred to the unitary
operations $\beta^{i}(t,x,y)$,
instead of the original $\tau^{i}$. Additionally, condition (C2)
will allow us to rewrite the QW evolution in terms of the usual state-dependent
translation operators. 

To alleviate the notations, in what follows we will omit the spacetime
dependence both in these matrices and in the $U_{i}(t,x,y)$, and
write simply $\beta^{i}$ and $U_{i}$. The above conditions allow
to calculate the $\beta^{i}$ matrices, which can be written as a
combination of Pauli matrices, i.e. $\beta^{i}=\vec{n}^{i}\cdot\vec{\sigma}$,
where each $\vec{n}^{i}$ must be a real, unit vector $\vec{n}^{i}=(\sin{\theta_{i}},0,\cos\theta_{i})$
for some angles $\theta_{i}$ (that are time and position
dependent).

In this way 
\begin{equation}
\beta_{i}=U_{i}^{\dagger}\sigma_{z}U_{i}=\left(\begin{matrix}\cos{\theta_{i}} &  & \sin{\theta_{i}}\\
\sin{\theta_{i}} &  & -\cos{\theta_{i}}
\end{matrix}\right),
\end{equation}
and each $U_{i}$ can be obtained by diagonalization of the corresponding
$\beta^{i}$. We finally write
them as 
\begin{equation}
U_{i}=\left(\begin{matrix}\cos{\frac{\theta_{i}}{2}} &  & \sin{\frac{\theta_{i}}{2}}\\
-\sin{\frac{\theta_{i}}{2}} &  & \cos{\frac{\theta_{i}}{2}},
\end{matrix}\right).\label{eq:choiceUis}
\end{equation}
The most naif way to implement such a walk is alternating the unitaries $U_i$ and the shift operators in the same manner we did in \ref{Th:DiracTrQW}, where the unitaries were homogeneous due to an appropriate choice of the coin state basis and its parameters. However, although different, the unitaries will be three different ones, one of them depending on a different parameter $\theta_i$, this strategy is not sufficient because $\Lambda$ depends on four real free parameters. 
Thus, in order to recover this local deformation we need at least one more internal parameter in the QW evolution. Moreover because we want to get terms like $\partial_j \lambda_{ij}$, as we have seen in the previous section, we have to iterate twice each unitary operator. All this considerations lead to build the following operator:
\begin{equation}
\Psi_{j+2,k}(r) = Z_2 Z_1 \Psi_{j,k}(r).
\end{equation}
where 
\begin{equation}
\begin{split}
Z_1 =  H \Pi^{2}_{i=0} \bar V_{i}V_{i}H\\
Z_2 =  Q \Pi^{2}_{i=0} \bar K_{i}K_{i}Q,
\end{split}
\end{equation}
and
\begin{equation}
\begin{split}
V_{i} = U_i S_{u_i} U^\dagger_i \\
\bar V_{i} = U^\dagger_i S_{u_i} U_i \\
K_{i}(\theta_{i+3}) = U_{i+3}S_{u_i} U_{i+3}^\dagger\\
\bar K_{i}(\theta_{i+3}) = U_{i+3}^\dagger S_{u_i} U_{i+3} 
\end{split}
\end{equation}
where
\begin{equation}
\begin{split}
H=\frac{1}{\sqrt{2}}\begin{pmatrix}1& 1\\1&-1\end{pmatrix} \hspace{1cm}
Q=\frac{1}{\sqrt{2}}\begin{pmatrix}1& -i \\1& i
\end{pmatrix}.
\end{split}
\end{equation}
Let us discuss this choice. Each $Z_i$ iterates twice the triangular QW as seen in \ref{Th:DiracTrQW}, choosing a different unitaries for each edge $k$ of the triangle in order to get in the continuous limit spatial derivatives of the unitaries $U_i$. Notice that for the second iteration we have chosen the conjugate transpose of the unitaries $U_i$. This is justified from the fact that, in the end we wish to recover spatial derivatives of the form $\partial_j \beta_i =(\partial_j U_i^\dagger)\sigma_z  U_i + U_i^\dagger \sigma_z \partial_j  U_i$. Iterating twice the same operator $V$ or $K$ would not be sufficient to recover the total derivative and we would have twice $U_i^\dagger \sigma_z \partial_j U_i$.  Finally $H$ and $Q$ ask for a change of basis in the coin state basis to recover in the continuous limit a true 2D propagation, as we already detailed in \ref{Th:DiracTrQW}. In conclusion, two of the $Z_i$ are necessary to have enough free parameters for the deformation $\Lambda$.
As in the previous section, we have chosen $\theta(r)_i = \pi + \varepsilon^{1/2} l_i(r)$ and $\Delta_x = \varepsilon^{1/2}$.

By expanding this equation up to first order in $\varepsilon$, after
a tedious but straightforward computation, which we spare the reader from detailing here, one arrives to the following
equation in the continuum limit: 

\begin{equation}\label{eq:continuumtriang}
\begin{split}
\partial_{t}\Psi&= \left[\partial_{u_2} (l_2 \sigma_x + l_5 \sigma_y ) +  \partial_{u_1} (l_1 \sigma_x + l_4 \sigma_y )+  \partial_{u_0}  (l_0 \sigma_x + l_3 \sigma_y )\right] \Psi+ \\
 +&\left[ (l_2 \sigma_x + l_5 \sigma_y )\partial_{u_2}  +  (l_1 \sigma_x + l_4 \sigma_y ) \partial_{u_1}  + (l_0 \sigma_x + l_3 \sigma_y )  \partial_{u_0}  \right]\Psi
\end{split}
\end{equation}

Notice that $\beta_i \simeq \lambda_i + o(\varepsilon)$. Now using the Eq. \ref{eq:shiftXY} we can reformulate the above equation in terms of $\partial_x$ and $\partial_y$: 

%\partial_2 = -\frac{1}{2}\partial_x-\frac{\sqrt{3}}{2}\partial_y
%\partial_1 = -\frac{1}{2}\partial_x+\frac{\sqrt{3}}{2}\partial_y
%\partial_0 = \partial_x

\begin{equation}\label{eq:continuumtriangXY}
\begin{split}
\partial_{t}\Psi&= \partial_x(\lambda_{00} \sigma_x + \lambda_{01} \sigma_y) \Psi +2 (\lambda_{00} \sigma_x + \lambda_{01} \sigma_y)  \partial_x\Psi\\
+&  \partial_y(\lambda_{10} \sigma_x + \lambda_{11} \sigma_y)  \Psi+2 (\lambda_{10} \sigma_x + \lambda_{11} \sigma_y)  \partial_y \Psi,
\end{split}
\end{equation}
where
\begin{equation}\label{eq:omega}
\begin{split}
\lambda_{00} = -\frac{1}{2}(l_2 + l_1) + l_0\\
\lambda_{01} = -\frac{1}{2}(l_5 + l_4) + l_3\\
\lambda_{10} = -\frac{\sqrt{3}}{2}(l_2- l_1) \\
\lambda_{11} = -\frac{\sqrt{3}}{2}(l_5 -l_4),
\end{split}
\end{equation}

which is a neat generalisation of the Eq. \ref{eq:diracmod} to an inohomogeneous $\Lambda$. Notice that we have a system of linear equations which is overdetermined and which leaves us enough freedom to recover the deformation matrix we wish. For instance a good choice to gauge away this ambiguity is: $l_5 = - l_4$ and $l_2 = - l_1$ which leads to the unique choice:
\begin{equation}\label{eq:omega}
\begin{split}
\lambda_{00} =  l_0 \hspace{0.5cm} \lambda_{01} = l_3 \hspace{0.5cm}  \lambda_{10} = \sqrt{3}l_1 \hspace{0.5cm}  \lambda_{11} = \sqrt{3}l_4 .
\end{split}
\end{equation}

We thus proved that a non homogeneous deformation of the triangulation can be simulated by local unitaries keeping the triangulation regular. 

\section{Discussion and perspectives}\label{sec:perspecitves}

In this chapter we proved that the QW hereby constructed over a general triangulation recovers, in the continuum limit, the Dirac equation
in curved $(2+1)$\textendash dimensional spacetime. We also noticed that the equation between changes of metric, aka $\Lambda-$deformations, and changes of local unitaries $\tau-$matrices enables us to absorb any continuous spatial deformations of the triangulation in the very fabric of the QW itself keeping the metrics flat (\ie paved by equilateral triangles). Besides being very powerful, this duality seems to be profound and deserves further development. It is worth mentioning that some physical theories provides a language to describe the quantum geometry of space using spin foam which are particular types of 2-complex, i.e. triangles. The principle of duality could therefore suggests new discrete symmetries underlying the continuous texture of space-time. Instead, from a technological point of view, the applications that the results presented in this chapter might have are many: for example, to mimic the transport on a carbon structure would mean to take into account its imperfections, the ripples, which would be simulated by a network of inhomogeneous local units, nowadays easily achievable with various technologies. But more generally any curved surface can be reproduced if the local deformation matrix is known. Other possible extensions are currently being developed: an extension to larger dimensions and the generalization to dynamic triangulations. An example of the latter is a very recent result where the triangulation is changed through Pachner moves, induced by the quantum walker density itself, allowing the surface to transform into any topologically equivalent one \cite{aristote2020dynamical}. This model extends the quantum walk over triangular lattice, introduced here. Unfortunately, if the triangulation is dynamical and there is interaction between such dynamics and the walker's dynamics, the equations will be highly non-linear and therefore analytically intractable. Finally, a recent result showed how the QW may be used to look for topological defects, which are properties of the configuration space itself \cite{roget2020grover}. We wonder if this suggests to target more general topological classification problems - \textit{e.g.} seeking to characterize homotopy equivalence over configuration spaces that represent manifolds as CW-complexes.

\newpage
 %17 pages -- proofread
% word count 4012

\chapter*{Conclusion}\label{chap:perspectives}

%toquote{quote }

\clearpage

\emph{Contributions.}~ In this thesis we reviewed the relevant Quantum Walks literature, and introduced all the necessary definitions for studying this unitary operators in one and higher dimensional space. We proved one important property about them, namely \textit{plasticity}, which unifies the two historical approach to quantum simulate transport equation, i.e. Hamiltonian simulation and digital quantum simulation.  We pointed out that in 2D plasticity allows to go beyond transport, allowing for dispersive terms in the PDE, i.e. have quadratic spatial derivatives. Coming back to the definition of QW, we extended their definitions to equilateral triangulations and finally we proved that this class of QW provides an elegant and general way to simulate transport equations over any arbitrary curved surface. 

\noindent \emph{Open questions.}~ As we presented the above results, we discussed several concrete open questions left to tackle. Because these were direct improvements on the results, it was easier to explain them `inline'. These included:
\begin{itemize}
\item Extending the plasticity theorem to higher dimensional cases than 2D and to any d-simplicial complex.
\item Investigating the duality principle and the retained symmetries of a QW over a general triangulation.
\item Seeking for a generalisation of the triangular QW in 3D spatial dimensions. 
\item Looking for a general definition of QWs coupled to reversible and causal dynamical triangulations. 
\item Investigating the question of how to properly generalise all the QW-based quantum simulation introduced in these theses to a multiparticle settings (namely QCA).
\item Seeking for fault tolerant QWs based quantum simulation model.
\end{itemize}

%\section{Modeling physical phenomena and transition to the continuum.}\label{sec:persp}

%\noindent \emph{From Simulation to Algorithmics.}  Quantum algorithms are generally considered to be a long-term applications of quantum computing. This is because of the common understanding that we will need to build scalable implementations of universal quantum gate sets with high fidelity first, and implement quantum error corrections then, in order to finally be able to run our preferred quantum algorithm on the thereby obtained universal quantum computer. This seems feasible, yet long way to go. A recent result of ours has proved that this may be a pessimistic view, proving that one of the most important algorithm, the Grover search, is in fact a naturally occurring phenomenon, e.g. when fermions propagate in crystalline materials under certain conditions. QWs are able to implement such mechanisms over a surface featuring topological—without the need for a specific oracle step. 
%
% We should investigate whether nature actually implements several others of these quantum algorithms ‘spontaneously’ allowing computer science to extract from nature its ability to run a computation. Using Plastic QW over general triangulation seems to be the most general framework for seeking such of them and once modeled, these quantum natural phenomena could be simulated efficiently by a quantum computer \\

\noindent \emph{Conclusion.}~ Looking back, the main lesson of this strand of research is that some local unitary operators in discrete space and discrete time, can simulate a wide range of transport phenomena. This is shown by computing the continuous limit and verifying that the leading orders converge, in the limit, to the physical laws we are seeking for. Conversely, we can think of some continuous physical evolutions as being the emergent/effective result of a microscopic discrete model. In such a view, nature may be grounded upon some very operational phenomena, in terms of operations over matrices and compositions of them. A detailed discussion of the logical relationship between physicality and computability at formal level may be found in \cite{ArrighiGANDY}.  Symmetries are essential in physics, i.e. all fundamental laws are justified as a byproduct of the invariance of the dynamics under a symmetry group. Therefore, it seems quite clear that investigating discrete symmetries in a discrete setting and their transition to the continuum is crucial if we want to establish quantum computation models as physical. 

\newpage
 %5 pages -- proofread

%-------------------------------------------------------------------------------------------
\backmatter
%------------------------------------------------------------------------------------------
\singlespace\pagestyle{plain}\pagenumbering{alph}

\bibliography{biblio}
\bibliographystyle{plain}

%\include{index}
%\addcontentsline{toc}{chapter}{Index}
\end{document}